\documentclass[fleqn,usenatbib]{mnras}

\usepackage{newtxtext,newtxmath}

\usepackage[T1]{fontenc}

\DeclareRobustCommand{\VAN}[3]{#2}
\let\VANthebibliography\thebibliography
\def\thebibliography{\DeclareRobustCommand{\VAN}[3]{##3}\VANthebibliography}

\usepackage{graphicx}	
\usepackage{amsmath}	

\usepackage{amssymb}	
\usepackage{multicol}
\usepackage{comment}
\usepackage{subcaption}

\newcommand{\mat}[1]{\boldsymbol{\mathsf{#1}}}

\title[Constraining growth rate with wide-angle effect]{Using peculiar velocity surveys to constrain the growth rate of structure with the wide-angle effect}

\author[Y. Lai et al.]{
Yan Lai,$^{1}$\thanks{E-mail: y.lai1@uqconnect.edu.au}
Cullan Howlett,$^{1}$
Tamara M. Davis$^{1}$
\\
$^{1}$School of Mathematics and Physics, The University of Queensland, QLD 4072, Australia\\
}

\date{Accepted XXX. Received YYY; in original form ZZZ}

\pubyear{2022}

\begin{document}
\label{firstpage}
\pagerange{\pageref{firstpage}--\pageref{lastpage}}
\maketitle

\begin{abstract}
Amongst the most popular explanations for dark energy are modified theories of gravity. The galaxy overdensity and peculiar velocity fields help us to constrain the growth rate of structure and distinguish different models of gravity. We introduce an improved method for constraining the growth rate of structure with the galaxy overdensity and peculiar velocity fields. This method reduces the modelling systematic error by accounting for the wide-angle effect and the zero-point calibration uncertainty during the modelling process. We also speed up the posterior sampling by around 30 times by first calculating the likelihood at a small number of fiducial points and then interpolating the likelihood values during MCMC sampling. We test the new method on mocks and we find it is able to recover the fiducial growth rate of structure. We applied our new method to the SDSS PV catalogue, which is the largest single peculiar velocity catalogue to date. Our constraint on the growth rate of structure is \(f\sigma_8 = 0.405_{-0.071}^{+0.076}\) (stat) \(\pm 0.009\) (sys) at the effective redshift of 0.073. Our constraint is consistent with a Planck 2018 cosmological model, \(f\sigma_8 = 0.448\), within one standard deviation. Our improved methodology will enable similar analysis on future data, with even larger sample sizes and covering larger angular areas on the sky. 

\end{abstract}




\section{Introduction}
Our current cosmological model, the \(\Lambda\) Cold Dark Matter model (\(\Lambda\)CDM) explains the accelerating expansion of the universe by introducing the cosmological constant into Einstein's theory of general relativity. Although this model is supported by numerous observations such as the Cosmic Microwave Background (CMB \citealt{Planck_2020}) and Type Ia supernovae \citep{Brout_2022}, the nature of dark energy is still unknown. An alternative explanation for the accelerating expansion is a modification to our theory of gravity (e.g., \citealt{Dvali_2000, De_Felice_2010}). 

The strength of gravity is different in different theories of gravity \citep{Linder_2007}. This affects the distribution of large-scale structures in the late universe and the motions of galaxies induced by these large-scale structures. The rate at which these structures grow is characterised by the linear growth rate parameter \(f(a) = \frac{d \ln{D}}{d \ln{a}}\), where \(a\) is the scale factor which describes the relative size of the universe at different epochs and \(D\) is the growth factor which describes how the matter overdensities grow in time. Hence, different theories of gravity will predict different linear growth rates at the same redshift. For example, the strength of gravity in the DGP model is weaker than the general relativity so it has a lower linear growth rate at the same redshift than general relativity \citep{Dvali_2000}. Some modified gravity theories also introduce a scale dependence to the growth rate. However, the differences among the linear growth rates in different theories of gravity may be small depending on the values of the additional degrees of freedom introduced by the model, so we require a high-precision measurement of the linear growth rate to confront these theories.

We can constrain the linear growth rate of structure either by directly measuring the peculiar velocities of galaxies or by quantifying the change in the galaxy distribution inferred from redshifts (which are contaminated by the galaxies' peculiar velocities). The second effect is called redshift space distortions (RSD; \citealt{Jackson_1972, Kaiser_1986}). 

The peculiar velocity of a galaxy is generated by its local gravitational interactions with other galaxies and is independent of the expansion of the universe. To measure the peculiar velocity, we need to first measure the total velocity of a galaxy spectroscopically. Then we can use the scaling relations such as the Tully-Fisher relation \citep{Tully_1977} for spiral galaxies or the fundamental plane \citep{Djorgovski_1987, Dressler_1987} for elliptical galaxies to determine the redshift-independent distance. Hubble's law allows us to use the redshift-independent distance to calculate the recession velocity due to the expansion of the universe. Finally, the difference between the total and recession velocity gives the peculiar velocity \citep{Davis_2014}. 

In linear theory, the peculiar velocity and galaxy density are only sensitive to the parameter combinations  \(b\sigma_8\) and \(f\sigma_8\), where \(\sigma_8\) is the root mean square of matter density fluctuation within spheres of radius \(8h^{-1} \mathrm{Mpc}\) and defines the overall normalisation of the density perturbations.\footnote{This degeneracy can be broken if we use the three-point correlation function/bispectrum or combine the result with weak lensing \citep{Gil_Mar_n_2015, Massey_2007}.} The peculiar velocity and RSD are highly complementary methods to measure the linear growth rate of structure, because while $b\sigma_8$ and \(f\sigma_8\) enter at the same order in equation~(\ref{eq:Kaiser_equation}), the peculiar velocity in equation~(\ref{eq:peculiar_velocity}) has no dependence on galaxy bias. The peculiar velocity is also more sensitive to the large-scale matter overdensity while the RSD is more sensitive on smaller scales \citep{Koda_2014}. Most importantly, the peculiar velocity and galaxy overdensity are two different tracers of the same underlying matter density field. Previous literature has shown combining two different tracers in the same analysis can eliminate the cosmic variance and reduce the statistical uncertainty \citep{McDonald_2009, Blake_2013, Koda_2014}. 

Several methods have been developed to combine both tracers to constrain the growth rate. For example, we can determine the growth rate of structure by measuring the two-point correlation functions \citep{Nusser_2017, Dupuy_2019, Turner_2021}, measuring the galaxy density and peculiar velocity fields and constraining the growth rate by maximizing the likelihood \citep{Johnson_2014, Huterer_2017, Howlett_2017, Adams_2017, Adams_2020}, measuring the momentum power spectrum \citep{Park_2000, Park_2006, Howlett_2019, Qin_2019} and using the density field to reconstruct the predicted peculiar velocity field \citep{Carrick_2015, Boruah_2020, Said_2020, lilow_2021}. In this work, we focus on the maximum likelihood fields method to constrain the growth rate of structure. 

In the development of the maximum likelihood fields method, \citet{Adams_2017} were the first to simultaneously model the auto- and cross-covariance matrices of galaxy overdensities and peculiar velocities and use these matrices to constrain \(f\sigma_8\) by maximizing the Gaussian likelihood function. They found the constraint on the linear growth rate of structure is improved by 20\% compared to analysis using a single tracer. \citet{Adams_2020} improved on this previous method by including the effect of RSD on the density field and velocity field. However, they assume the lines-of-sight to all galaxies are parallel to each other during the derivation. This assumption breaks down when the angle subtended by the lines-of-sight between two different galaxies is big and the error introduced is called the wide-angle effect. This wide-angle effect is expected to be particularly important in a peculiar velocity survey because they are usually much shallower and wider than galaxy surveys at higher redshift. Previous literature has shown the wide-angle effect can introduce a more than 10\% error when the subtended angle between the two line-of-sights is greater than 30 degrees \citep{Castorina_2018, Maresuke_2021}. 

\citet{Castorina_2020} recently developed a wide-angle formalism for the galaxy overdensity and peculiar velocity two-point correlation functions, which we build on to include wide-angle effects in the maximum likelihood field method from \citet{Adams_2020}. \citet{Castorina_2020} did not consider the damping in the galaxy and velocity power spectrum due to the effect of RSD on small scales (finger-of-god effect) \citep{Koda_2014}, so we also extend their methodology to include these non-linearities. In essence, our study combines the previous work by \citet{Adams_2020} and \citet{Castorina_2020} to develop a new maximum likelihood method that considers both the wide-angle effect and the RSD. As we will show, this reduces the systematic error when constraining the growth rate of structure from combined redshift and peculiar velocity surveys, while maintaining the same constraining power as the method of \citet{Adams_2020}. We apply our new method to obtain the first cosmological constraints from the newly released Sloan Digital Sky Survey peculiar velocity (SDSS PV) catalogue, which is the largest individual sample of peculiar velocities to date.

This paper is organized as follows. In Section 2, we introduce the SDSS data and simulated catalogues. In Section 3, we introduce the theory behind our new method and derive analytical formulae for the models. The wide-angle effect, and a number of additional considerations that are required to analyse the SDSS PV catalogue, will be introduced in Section 4. We will show the result of tests of our methodology on the SDSS mocks in Section 5. Section 6 then presents the constraint of the growth rate of structure from the SDSS PV catalogue and discusses its implications and comparison to previous constraints. Lastly, we present our conclusion in Section 7. Throughout this work, we assume a flat \(\Lambda\)CDM model with fiducial cosmological parameters given by \(\Omega_m = 0.3121\), \(H_0 = 100h \mathrm{km} \mathrm{s}^{-1} \mathrm{Mpc}^{-1} \) and \(\sigma_8 = 0.8150\) at redshift zero.  

\section{Data and simulation}
\subsection{SDSS peculiar velocity catalogue}
\begin{figure}
	\includegraphics[width=\columnwidth]{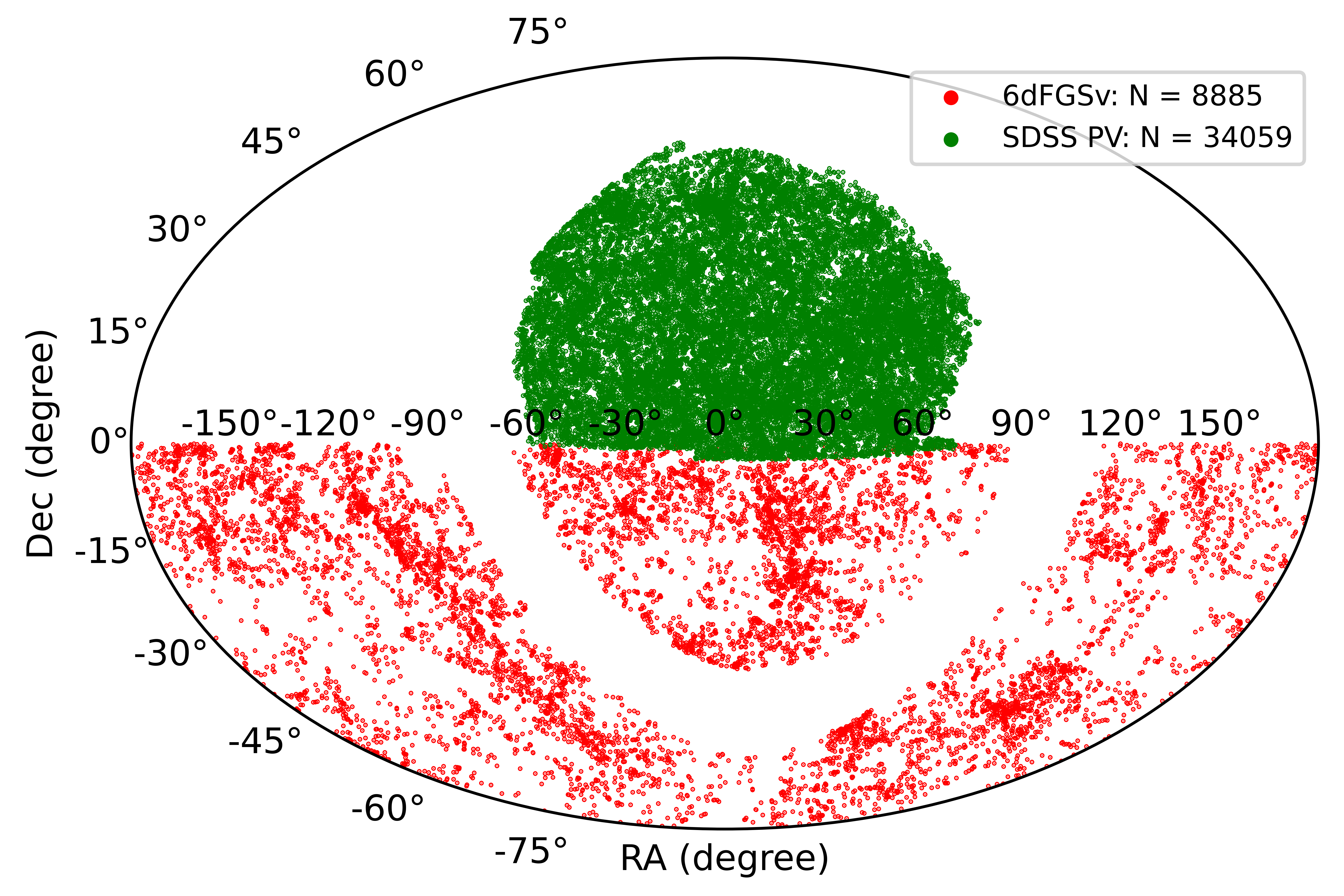}
    \caption{This plot shows the distribution of galaxies in 6dFGSv and SDSS PV catalogue in the Aitoff projections. The 6dFGSv covers the whole southern hemisphere except the zone-of-avoidance region while the SDSS PV sample covers a small patch of the northern hemisphere. Despite containing fewer galaxies than the SDSS PV sample, the area covered by the 6dFGSv sample is around 3 times larger than the SDSS PV sample. }
    \label{fig:Distribution}
\end{figure}

\begin{figure}
	\includegraphics[width=\columnwidth]{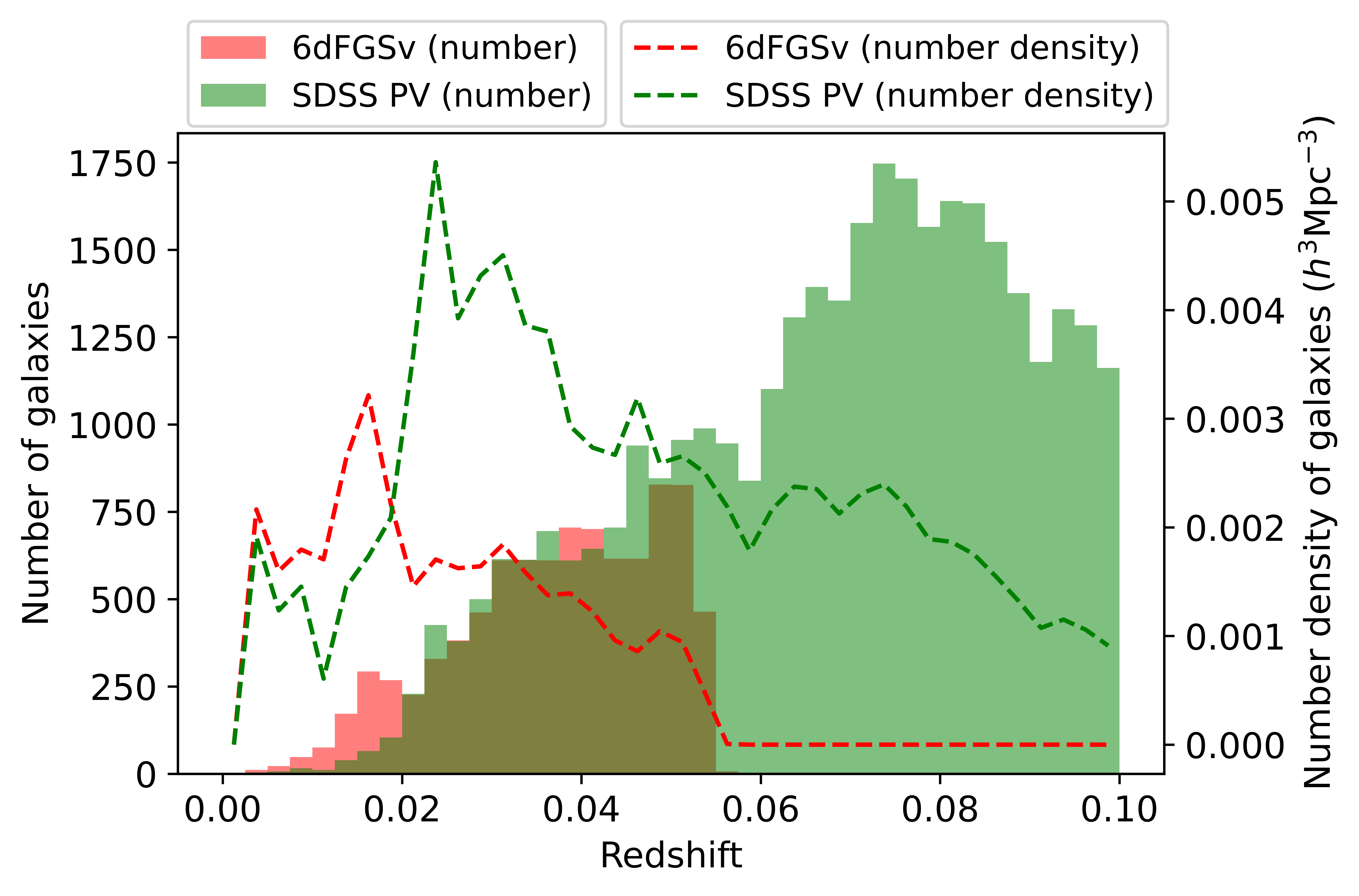}
    \caption{This plot shows the redshift distribution of galaxies in the 6dFGSv and SDSS PV catalogue. The histogram shows the number of galaxies in each redshift bin and the dotted line shows the number density of galaxies in each redshift bin. This plot demonstrates most of the galaxies from the SDSS PV sample come from a redshift range above 0.05 which is above the redshift limit for the 6dFGSv sample. Below the redshift of 0.02, the 6dFGSv sample contains more galaxies. Above the redshift of 0.02, SDSS PV sample typically contains more galaxies than the 6dFGSv sample.}
    \label{fig:redshift_distribution}
\end{figure}

The SDSS PV catalogue contains 34,059 elliptical galaxies with distance and peculiar velocity measurements obtained from the fundamental plane relation \citep{Howlett_2022}. The targets in the SDSS PV catalogue were selected from data release (DR) 14 of SDSS and it covers 7016 \(\mathrm{deg}^2\) in the northern hemisphere up to redshift of 0.1 with a mean uncertainty in log-distance ratio of 0.1 dex. The SDSS PV sample is around four times bigger than the previous biggest fundamental plane peculiar velocity survey -- the 6-degree Field Galaxy Survey velocity sample (6dFGSv; \citealt{Campbell_2014}). Fig.~\ref{fig:Distribution} illustrates the 6dFGSv sample covers an area around 3 times bigger than the SDSS PV sample. Fig.~\ref{fig:redshift_distribution} demonstrates the SDSS PV sample is much deeper than 6dFGSv sample. Additionally, the number of galaxies in the SDSS PV sample in each redshift bin is higher than 6dFGSv except below the redshift of 0.02. The SDSS PV sample extends to a higher redshift than other current large peculiar velocity samples. As such, it allows probing of slightly larger cosmological volumes. At the same time, the uncertainties of peculiar velocities for the high redshift galaxies are bigger, so more galaxies are required to obtain a measurement of similar accuracy as the low redshift galaxies. 

\begin{figure}
	\includegraphics[width=\columnwidth]{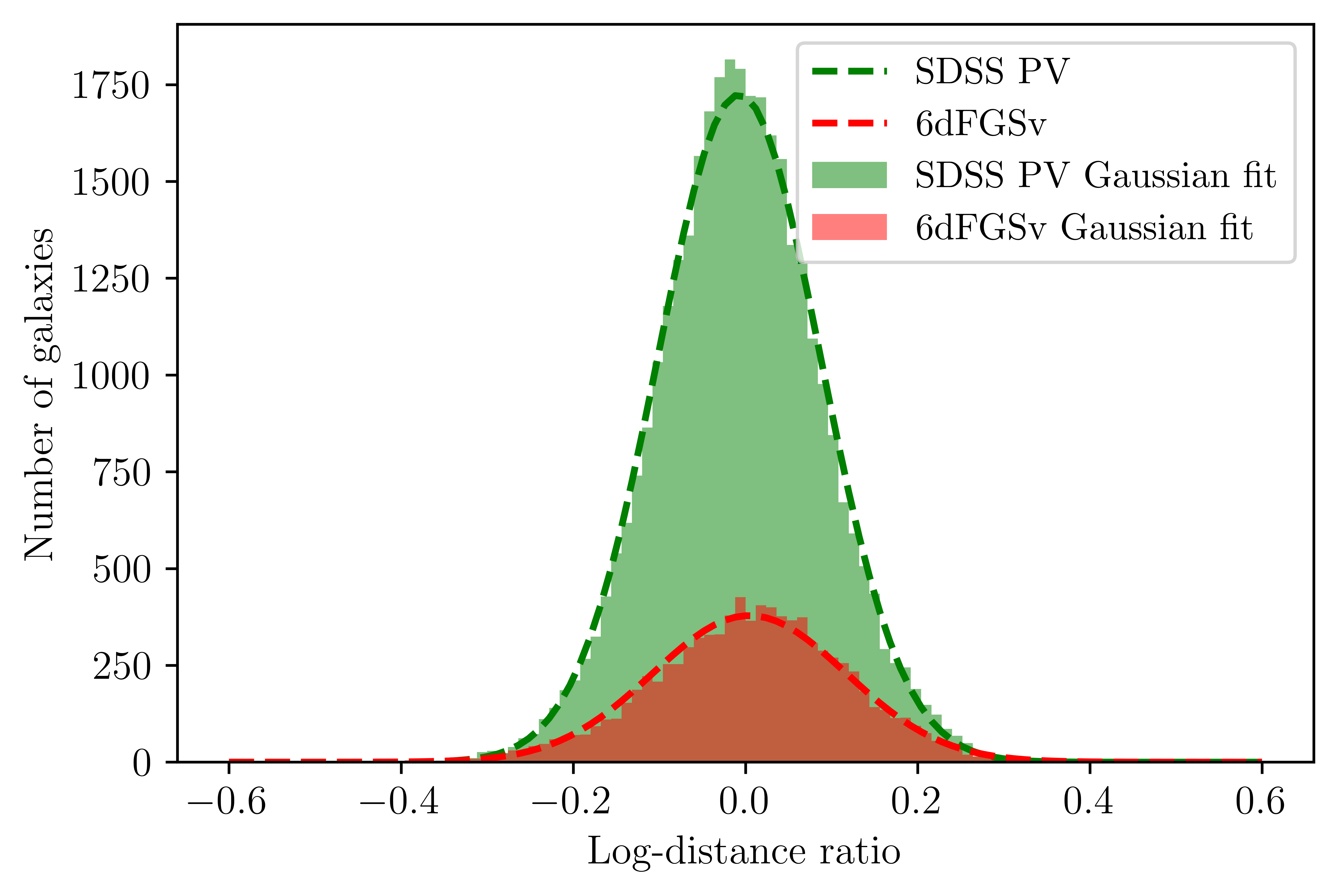}
    \caption{This plot shows the distribution of the log-distance ratios in the 6dFGSv and SDSS PV sample. The log-distance ratio is defined in equation~(\ref{eq:log-dist}) and is analogous to peculiar velocity. This figure shows the log-distance ratio in both SDSS PV and 6dFGSv samples follow roughly the Gaussian distribution with zero mean as expected.}
    \label{fig:log_dist_ratio}
\end{figure}

\begin{figure*}
\begin{multicols}{2}
    \includegraphics[width=\linewidth]{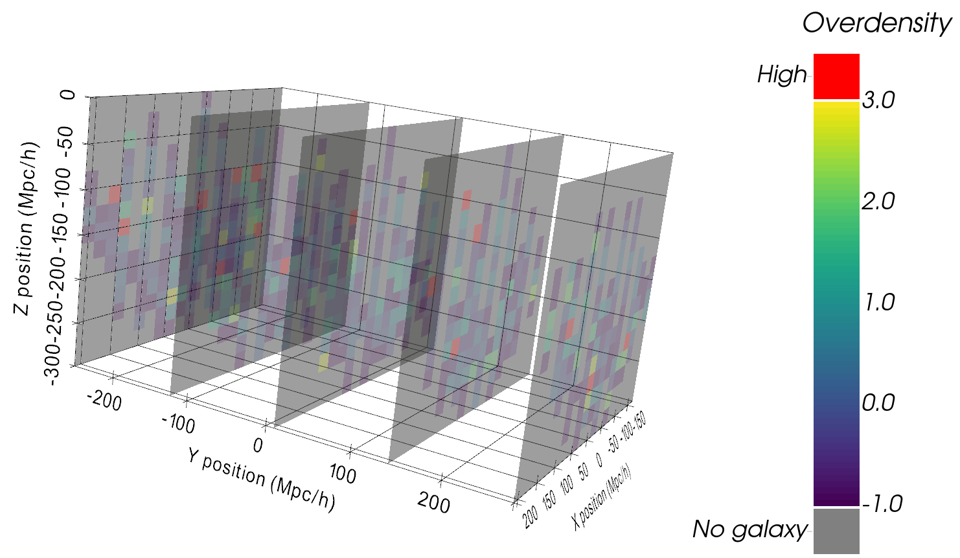}\par 
    \includegraphics[width=\linewidth]{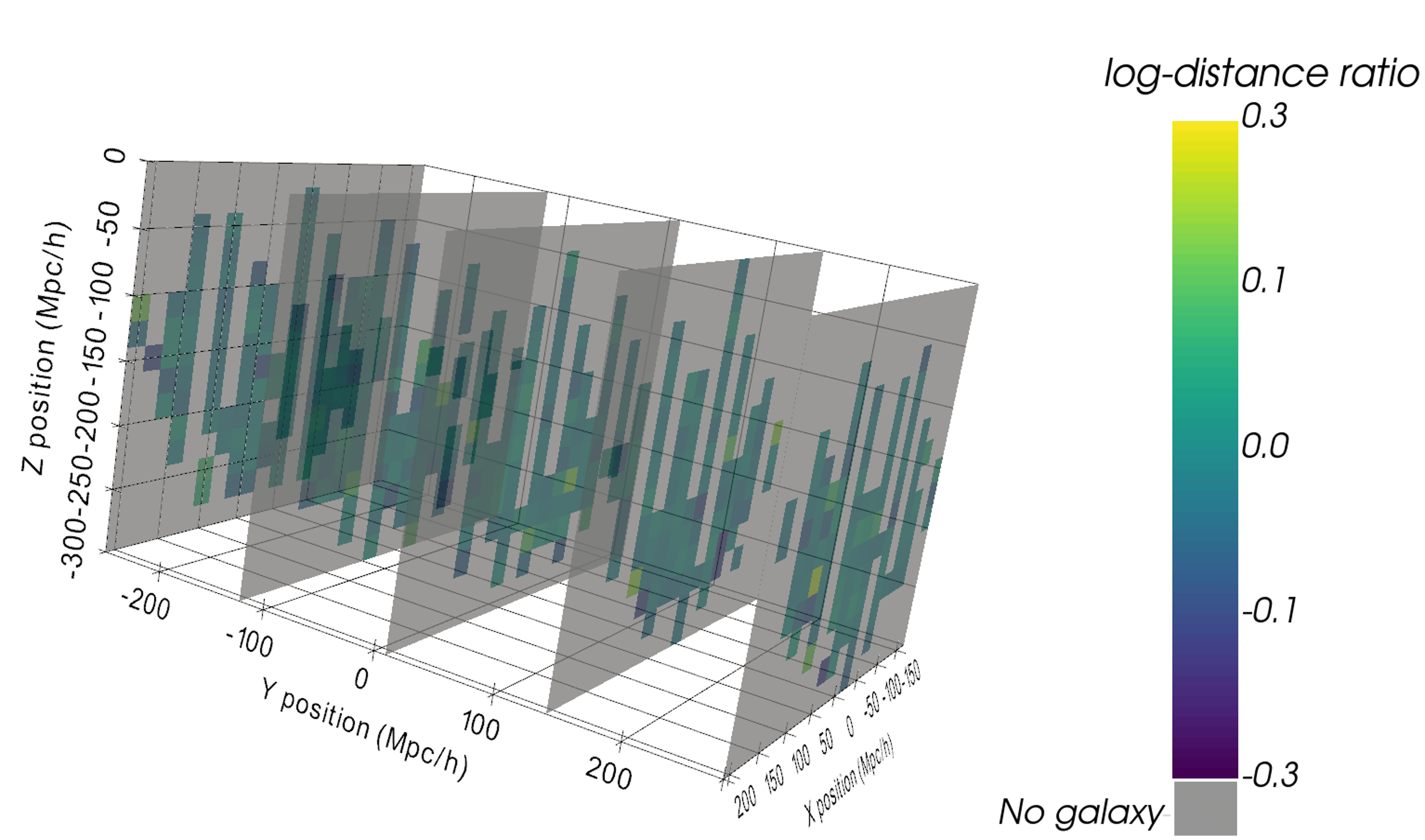}\par
\end{multicols}
\caption{These plots show the slices through the SDSS PV catalogue in the y-axis for the galaxy overdensity (left) and the log-distance ratio (right). The observer is at the origin and the slices are at -260 \(h^{-1} \mathrm{Mpc}\), -125 \(h^{-1} \mathrm{Mpc}\), 10 \(h^{-1} \mathrm{Mpc}\), 145 \(h^{-1} \mathrm{Mpc}\) and 280 \(h^{-1} \mathrm{Mpc}\) in the y-direction relative to the observer. These data are arranged in cubic grid cells with width 20 \(h^{-1} \mathrm{Mpc}\) (see section \ref{sec:gridding}). Left panel: the grey transparent regions indicate there is no galaxy in that grid cell and the red regions indicate regions with galaxy overdensity higher than three. We can see \(\delta_g \lesssim 3.0\) for most of the grid cells, so the quasi-linear theory model can accurately describe our galaxy overdensity data. Right panel: the grey transparent regions indicate no velocity measurement in that grid cell. We can see most of the log-distance ratios are very close to zero which agrees with Fig~\ref{fig:log_dist_ratio}. If we convert the log-distance ratio measurements here to the peculiar velocity, it shows the average peculiar velocity in each grid cell is around a few hundred to a few thousand km/s. }
\label{fig:slice_plot}
\end{figure*}

Fig.~\ref{fig:log_dist_ratio} shows the distribution of log-distance ratios (defined in equation~\ref{eq:log-dist}) is roughly Gaussian with zero mean as expected. In Fig.~\ref{fig:slice_plot}, we plot the galaxy overdensity and log-distance ratio in slices of the SDSS PV catalogue in the y-direction. We find \(\delta_g \lesssim 3.0\) for most of the data, so the quasi-linear theory in \citet{Koda_2014} would produce a reasonably accurate model for our data. Most of the log-distance ratio data is also very close to zero which agrees with Fig.~\ref{fig:log_dist_ratio}. If we convert the log-distance ratio measurements to peculiar velocity, the typical peculiar velocities of galaxies in the SDSS PV catalogue are around a few hundred to a few thousand km/s. For a detailed discussion of the data selection process, we refer the readers to section 2 of \citet{Howlett_2022}. 

After fitting the fundamental plane and log-distance ratio, \citet{Howlett_2022} found the peculiar velocity depends on the richness of the galaxy group with smaller peculiar velocities when the group contains more galaxies. The origin of this bias is unknown and it was corrected by fitting different Fundamental plane for different group richness. In this work, we will use the peculiar velocity data from the catalogue of multiple Fundamental plane fits instead of a single fit. 

One of the assumptions of the fundamental plane fit is that the net velocity of all galaxies inside the sample is zero. The zero-point correction corrects this assumption. For the SDSS PV catalogue, the zero-point assumption is done by cross-matching the overlapping galaxy groups in the Cosmicflows-III catalogue \citep{Tully_2016}. The mean difference between the log-distance ratios of these overlapping galaxies in the cosmicflow-III and SDSS PV sample gives the zero-point correction. \citet{Howlett_2022} calculates the zero-point correction for both single and multiple fundamental plane fit, we will adopt the value from the multiple fundamental plane fit and also propagate the uncertainty of this callibration into our cosmological constraints. 

\subsection{Mock catalogues}
The mock peculiar velocity catalogues are produced to closely replicate the data of the SDSS PV catalogue and allow us to test whether our methodology is unbiased before it is applied to the data. The 2048 mocks are produced with 256 approximate N-body dark matter \textsc{L-PICOLA} simulations \citep{Howlett_2017} with a flat \(\Lambda\)CDM cosmological model with the fiducial cosmology at redshift zero. These input cosmological parameters give the fiducial value of \(f\sigma_8 = 0.432\) at redshift zero. Each simulation is a cubic box containing \(2560^3\) particles evolved to redshift zero with a side length of \(1800h^{-1} \mathrm{Mpc}\) \citep{Howlett_2022}. Each simulation produces eight different mocks by placing eight different observers at different locations in the simulation. To minimise the correlation between different mocks in the same simulation, the observers are placed at least \(600h^{-1} \mathrm{Mpc}\) apart. 


Once we populate the galaxies inside the simulation, the angular mask is applied to the mocks to ensure the mock catalogues match the footprint of the data. The perturbed fundamental plane parameters are generated for all galaxies with a Gaussian random number generator centered on the truth value. The apparent magnitude is then found based on the fundamental plane parameters. Lastly, the selection function of the data is applied to ensure the mocks match the data \citep{Howlett_2022}. For a more detailed description of the mock generation, we refer the readers to section 3 of \citet{Howlett_2022}.  

After fitting the Fundamental plane with the mocks, \citet{Howlett_2022} discovered the observed peculiar velocities in the mocks were correlated with the absolute magnitudes of the galaxies. However, they found this correlation is the by-product of using the Fundamental plane as a distance indicator and will not introduce bias in cosmological constraints. 

\section{Theoretical modelling}
In this section, we present an overview of the new model we use to fit the positions and velocities of the SDSS PV catalogue as a function of cosmological parameters. The inclusion of wide-angle effects, redshift-space distortions, and various methods to improve the comparison to data (such as gridding and marginalization over zero-point errors) make the modelling quite complex, so we leave detailed derivations of many of the expressions to the appendices. We start with an overview of our geometry and notation.
\subsection{Geometry and notation}
\begin{figure}
    \centering
	\includegraphics[width=0.4\columnwidth]{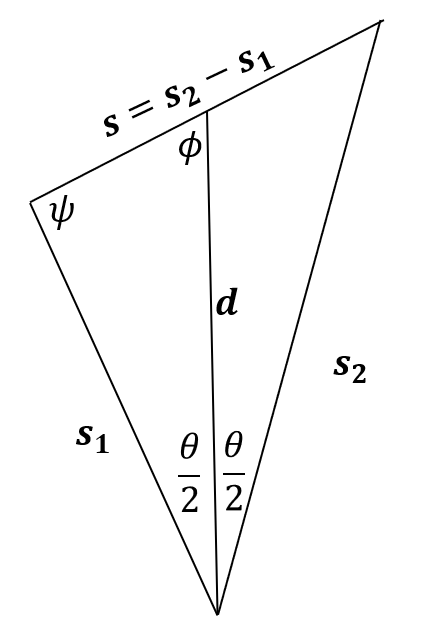}
    \caption{This plot shows the geometry among the observers and two random galaxies. The vectors \(\boldsymbol{s_1}\) and \(\boldsymbol{s_2}\) denote the radial distance between the observer and the first/second galaxy. The distance between the two galaxies is represented by \(\boldsymbol{s}\). The angular bisector of the two galaxies is denoted by \(\boldsymbol{d}\) and the angle subtended by the two galaxies is \(\theta\). Lastly, \(\phi\) represents the angle between the angular bisector and the separation vector of the two galaxies.}
    \label{fig:Geometry}
\end{figure}
Fig.~\ref{fig:Geometry} shows the configuration of two random galaxies at distance \(\bold{s_1}\) and \(\bold{s_2}\) relative to the observer.\footnote{Throughout this paper, we use bold letters to indicate a vector quantity and \(\hat{n}\) to represent the unit vector of \(\bold{n}\).} The distance between the two galaxies is denoted by \(\bold{s} = \bold{s_2} - \bold{s_1}\) and the angular bisector of these two galaxies is given by 
\begin{equation}
    \bold{d} = \frac{s_1 s_2}{s_1 + s_2} (\hat{s_1} + \hat{s_2}).
    \label{eq:distance}
\end{equation}
We can calculate the angle subtended by the two galaxies \(\theta\) by taking the dot product of \(\hat{s_1}\) and \(\hat{s_2}\). Similarly, the angle \(\phi\) is computed by taking the dot product of \(\bold{s}\) and \(\bold{d}\). In this latter definition, there are two edge cases that must be given special consideration because $\phi$ becomes undefined. These are when two galaxies are directly opposite each other along the line-of-sight (\(\hat{s_1} = - \hat{s_2}\)), and when two galaxies are at the same location \(\bold{s_1} = \bold{s_2}\). In the first of these scenarios, we can use the angular bisector theorem (Eq.~\ref{eq:phi}) to show \(\phi = \frac{\pi}{2}\). The second is irrelevant because our derivation shows in this case, our model only ever depends on \(\theta\) and not \(\phi\).


\subsection{The maximum likelihood fields method}
In the maximum likelihood fields method, we constrain the linear growth rate by maximising the likelihood of observing a set of galaxy overdensities and peculiar velocities. 


The large-scale structures of our Universe are believed to have formed from the primordial Gaussian fluctuations in the density field \citep{Zeldovich_1970}. On linear scales, the velocity fluctuations are also expected to follow the density fluctuation. Therefore, on large enough scales, we can model the likelihood function with a Gaussian distribution
\begin{equation}
    P(\boldsymbol{S}|\boldsymbol{m}) = \frac{1}{\sqrt{(2 \pi)^n |\mat{C(m)}|}} e^{-\frac{1}{2} \boldsymbol{S}^T \mat{C}(\boldsymbol{m})^{-1}\boldsymbol{S}},
    \label{eq:Gaussian_like}
\end{equation}
where \(n\) is the length of the data vector, \(\mat{C(m)}\) is the covariance matrix which depends on the model parameters $\boldsymbol{m}$, and \(\boldsymbol{S} = (\boldsymbol{\delta}_{g},\boldsymbol{v})\) denotes the data vector that contains the galaxy \textit{over}density and peculiar velocity information. In defining our position and velocity information in terms of the overdensity and peculiar velocity we can assume the mean values of these two fields are zero based on the cosmological principle. Therefore, they do not enter into the likelihood function (note that this assumption may not be exactly true for peculiar velocity data in practice due to the presence of zero-point systematics, which we additionally model in Section~\ref{sec:zeropoint}).

In order to maximise the likelihood given our SDSS data, we hence require a model for the covariance matrix of the observed overdensity and peculiar velocity fields as a function of cosmological and nuisance parameters. The full covariance matrix is decomposed into blocks as 
\begin{equation}
\boldsymbol{\mathsf{C}} = 
\begin{pmatrix}
    \mat{C}_{gg} & \mat{C}_{gv} \\
    \mat{C}_{vg} & \mat{C}_{vv}
\end{pmatrix}
.
\label{eq:full_bf}
\end{equation}
where $\mat{C}_{gg}$ and $\mat{C}_{vv}$ denote auto-covariance matrices of the galaxy overdensity and velocity fields respectively, while $\mat{C}_{gv}$ is the cross-covariance matrix. The full covariance matrix allows us to constrain the linear growth rate of structure with galaxy overdensity and peculiar velocity simultaneously. This will help to reduce the uncertainty of the linear growth rate of structure by breaking the cosmic variance limit because galaxy density and peculiar velocity are two different tracers of the same underlying matter density field \citep{McDonald_2008}. Our models for the individual blocks are presented next.


\subsection{Covariance matrices}
In linear theory, the peculiar velocity \(\boldsymbol{v}\) of a galaxy is linked to the matter overdensity through \citep{Strauss_1995}
\begin{equation}
    \nabla \cdot \boldsymbol{v} = - a H f \theta_v = - a H f \delta_m,
    \label{eq:peculiar_velocity}
\end{equation}
where the Hubble parameter is denoted by \(H\) and the matter overdensity is represented by \(\delta_m\). The velocity divergence field is denoted by \(\theta_v\). The second equality is only true in the linear regime. We cannot observe matter overdensity directly because most of the matter in the universe is dark matter. Instead, we can only observe the galaxy overdensity. Additionally, we measure the galaxy distribution in redshift space, so we have to take into account the RSD effect. \citet{Kaiser_1986} found the galaxy overdensity in redshift space is given by 
\begin{equation}
    \delta_g^s = b \delta_m^r + f \mu^2 \theta_v.
    \label{eq:Kaiser_equation}
\end{equation}
The Kaiser equation links the galaxy overdensity in redshift space \(\delta_g^s\) to the matter density in real space \(\delta_m^r\) and the velocity divergence field by the galaxy bias parameter \(b\), the cosine of the line-of-sight angle \(\mu\) and the linear growth rate of structure \(f\). 

In the mildly non-linear regime, the galaxy overdensity can be modelled by extending equation~(\ref{eq:Kaiser_equation}) as
\begin{equation}
    \delta_g(\boldsymbol{k}) = \left(b \delta_m(\boldsymbol{k}) + f \mu^2 \theta_{v}(\boldsymbol{k})\right) D_g (k, \sigma_g, \mu),
\end{equation}
where \(D_g\) is the damping in the galaxy overdensity due to the finger-of-god effect \citep{Peacock_1994}. In this paper, we will adopt the Gaussian parameterization of the finger-of-god effect \citep{Adams_2017, Adams_2020}
\begin{equation}
    D_g = e^{-\frac{(k\mu\sigma_g)^2}{2}}.
\end{equation}
Here, \(\sigma_g\) determines the strength of damping in the unit of \(h^{-1} \mathrm{Mpc}\). In general, the cosine of the line-of-sight angle is given by \(\mu_{i} = \hat{s_{i}} \cdot \hat{k}\), where \(\hat{s_{i}}\) is the line-of-sight direction to galaxy $i$ and \(\hat{k}\) is the unit vector for wavenumber \(\bold{k}\). The Lorentzian parameterization of the finger of god effect \citep{Dekel_1999, Taylor_2001, Burkey_2004, Howlett_2017_b} is also popular in the literature. We will show later our method can be easily applied to the Lorentzian parameterization as well. The definition of a covariance matrix \(\mat{C}_{XY}\) of two arbitrary quantities \(X\) and \(Y\) is \begin{equation}
    \mat{C}_{XY} = \langle XY^* \rangle - \langle X \rangle \langle Y \rangle. 
    \label{eq:cov}
\end{equation}
The second term on the right-hand side of equation~(\ref{eq:cov}) vanishes because the mean overdensity and peculiar velocity are zero according to the cosmological principle. By applying equation~(\ref{eq:cov}) and considering the geometry in Fig.~\ref{fig:Geometry} we can find the galaxy overdensity auto-covariance matrix in Fourier space is
\begin{equation}
\begin{split}
        \mat{C}_{gg} = \int \frac{d^3 k}{(2 \pi)^3} e^{i \bold{k} \cdot \bold{s_1}}\int \frac{d^3 k'}{(2 \pi)^3} e^{-i \bold{k'} \cdot \bold{s_2}} \langle \delta_g(\bold{k}) \delta_g (\bold{k'})^* \rangle \\ 
        = \int \frac{d^3 k}{(2 \pi)^3} e^{i \bold{k} \cdot \bold{s}} (b^2 P_{mm} + bf \mu_1^2 P_{m\theta} +  bf \mu_2^2 P_{m\theta} + \\
        f^2 \mu_1^2 \mu_2^2 P_{\theta \theta}) e^{-\frac{(k \mu_1 \sigma_g)^2}{2}} e^{-\frac{(k\mu_2\sigma_g)^2}{2}}.
\end{split}
\label{eq:gg_be}
\end{equation}
Here, the matter auto-power spectrum, the cross-power spectrum of matter, and velocity divergence and the velocity divergence auto-power spectrum are denoted by \(P_{mm}\), \(P_{m\theta}\) and \(P_{\theta \theta}\) respectively. We use the definition of the power spectrum \(\langle X(k) Y(k')^* \rangle = (2 \pi)^3 \delta_{D} (k-k') P_{XY}(k)\) to simplify equation~(\ref{eq:gg_be}) to the second line.\footnote{\(\delta_{D}(k-k')\) here denotes the Dirac delta function, not to be confused with galaxy or matter overdensity.}

Similarly, the model for the peculiar velocity in the linear theory is 
\begin{equation}
    \boldsymbol{v}(\boldsymbol{k}) = - i a H f \frac{\mu}{\bold{k}} \theta_{v}(k) D_u(k, \sigma_u)
    \label{eq:v_model}
\end{equation}
in the Fourier space. Here, \(D_u = \frac{\sin(k\sigma_u)}{k\sigma_u}\) is the damping function for the peculiar velocity field due to the finger-of-god effect \citep{Koda_2014}. The \(\sigma_u\) parameter determines the strength of the damping. Substituting this equation into the definition of the velocity auto-covariance matrix, we get 
\begin{equation}
    \begin{split}
        \mat{C}_{vv} = \int \frac{d^3 k}{(2 \pi)^3} e^{i \bold{k} \cdot \bold{s_1}}\int \frac{d^3 k'}{(2 \pi)^3} e^{-i \bold{k'} \cdot \bold{s_2}} \langle \boldsymbol{v}(\bold{k_1}) \boldsymbol{v}(\bold{k_2})^* \rangle \\
        = \int \frac{d^3 k}{(2 \pi)^3} e^{i \bold{k} \cdot \bold{s}} \frac{(aHf)^2}{k^2} D_u^2 (k, \sigma_u) \mu_1 \mu_2 P_{\theta \theta}.
    \end{split}
    \label{eq:vv_be}
\end{equation}
We can also show the galaxy-velocity cross-covariance matrix and the velocity-galaxy cross-covariance matrix are given by 
\begin{equation}
    \begin{split}
        \mat{C}_{gv} = i a H f \int \frac{d^3 k}{(2 \pi)^3} e^{i \bold{k} \cdot \bold{s}} \frac{\mu_2}{k} D_u(k, \sigma_u) \\
        (b P_{m\theta} + f \mu_1^2 P_{\theta \theta})D_g(k, \sigma_g, \mu_1) 
    \end{split}
    \label{eq:gv_be}
\end{equation}
and 
\begin{equation}
    \begin{split}
        \mat{C}_{vg} = - i a H f \int \frac{d^3 k}{(2 \pi)^3} e^{i \bold{k} \cdot \bold{s}} \frac{\mu_1}{k} D_u(k, \sigma_u) \\
        (b P_{m\theta} + f \mu_2^2 P_{\theta \theta})D_g(k, \sigma_g, \mu_2) 
    \end{split}
    \label{eq:vg_be}
\end{equation}
respectively. 

\citet{Adams_2020} simplified the integral by setting \(\mu_1 = \mu_2 = \hat{k} \cdot \hat{d}\) while \citet{Castorina_2020} did not include the damping terms \(D_g\) and \(D_u\). To avoid these two simplifications, we first use the Taylor expansion to write the \(D_g\) damping term as 
\begin{equation}
    D_g = e^{-\frac{(k \mu \sigma_g)^2}{2}} = \sum_{i = 0}^{\infty} \frac{(-1)^i (k\sigma_g)^{2i}}{2^i i!} \mu^{2i}. 
    \label{eq:D_g_Taylor}
\end{equation}
The radius of convergence of this Taylor expansion is infinite, so equation~(\ref{eq:D_g_Taylor}) is exact if we sum to infinity. The Lorentzian parametrization of \(D_g\) could also be decomposed into a similar sum but with different coefficients using the Taylor expansion. We can then substitute this into equations~(\ref{eq:gg_be}-\ref{eq:vg_be}) and simplify the expressions into a series of much faster 1D integrals. The detailed derivations of the covariance matrices are shown in Appendix~\ref{sec:appendix_A} and we will show the final equations after simplifications here. The galaxy auto-covariance matrix is given by 
\begin{align}
        & \mat{C}_{gg} (\bold{s_1}, \bold{s_2}) = \sum_{p q} \frac{(-1)^{p+q}}{2^{p+q} p! q!} \sigma_g^{2(p+q)}
        \sum_{l} i^l   \biggl(b^2 \xi_{mm, l}^{p, q, 0}(s, 0) \notag \\
        & H_{p, q}^{l}(\bold{s_1}, \bold{s_2})+ f^2 \xi_{\theta \theta, l}^{p, q, 0}(s,0) H_{p+1, q+1}^{l}(\bold{s_1}, \bold{s_2})+ \notag \\
        & bf \xi_{m\theta, l}^{p, q, 0}(s, 0)\biggl[H_{p+1, q}^{l}(\bold{s_1}, \bold{s_2}) + H_{p, q+1}^{l}(\bold{s_1}, \bold{s_2})\biggl]\biggl).
    \label{eq:gg_af}
\end{align}
Like equation~(\ref{eq:D_g_Taylor}), this equation is exact if we sum \(p, q\) to infinity. In practice, we can only sum a finite number of terms. \citet{Adams_2020} points out the order of \(l\) only depends on the exponent of \(\mu\) that appears in the anisotropic power spectrum. The galaxy auto-covariance matrix has two damping terms. From equation~(\ref{eq:D_g_Taylor}) and equation~(\ref{eq:gg_be}), the highest possible exponent is \(2(p+q)\), so \(l \leq 2(p+q)\). The highest \(p,q\) we need to sum to depends on the required accuracy of the Taylor expansion. From equation~(\ref{eq:D_g_Taylor}), the accuracy of the Taylor expansion depends on \(k\sigma_g\). \citet{Koda_2014} found the model we are using is valid for \(k \leq 0.20h \mathrm{Mpc}^{-1}\) and \(\sigma_g\) is on the order of \(1 h^{-1} \mathrm{Mpc}\). Therefore, \(k\sigma_g \lesssim 1\) and the first few orders of Taylor expansion will give a reasonable approximation. In this work, we consider the Taylor expansion up to the third order (\(p=q=3\)) and show it is enough to obtain an unbiased constraint on \(f\sigma_8\) for the SDSS PV survey. \footnote{You can calculate the expressions for the covariance matrix with a higher order of Taylor expansion with the notebook we provide. However, increasing \(p, q\) will also significantly increase the computational time for the covariance matrix. The number of terms you need to calculate is given by \(2(p+q+2)\) for the galaxy auto-covariance matrix, so for \(p = q = 3\), you need 16 terms in total. It takes us about 30 hours to compute all the necessary galaxy auto-covariance matrix components on a single core.} 

The function \(\xi\) is given by 
\begin{equation}
    \xi_{ab, l}^{p, q, n}(r, \sigma_u) = \int_{k_{\mathrm{min}}}^{k_{\mathrm{max}}} \frac{k^2 dk}{2 \pi^2} P_{a b}(k) j_l(kr) k^{2(p+q)} D_u^n(k, \sigma_u),
    \label{eq:xi}
\end{equation}
where \(j_l\) is the \(l^{\mathrm{th}}\) order of the spherical Bessel function. For the galaxy auto-covariance matrix, if we set \(l = 0\), then \(\xi\) represents the galaxy two-point correlation function. Mathematically, the integration for equation~(\ref{eq:xi}) should be done from \(k = 0\) to \(k = \infty\) because equation~(\ref{eq:xi}) is a Fourier transform. However, a peculiar velocity survey can only probe a limited amount of $k$-modes. Therefore, the integration bound is set to \(k_{\mathrm{min}}\) and \(k_{\mathrm{max}}\) which are the largest and smallest scales that the peculiar velocity survey is able to probe respectively. For the SDSS peculiar velocity catalogue, we set \(k_{\mathrm{min}} = 0.0025h \mathrm{Mpc}^{-1}\) because this is approximately the largest scale that the survey is able to probe. \(k_{\mathrm{max}}\) depends on the accuracy of our model on small scales. Previous literature usually set \(k_{\mathrm{max}} = 0.15h \mathrm{Mpc}^{-1}\) \citep{Adams_2017, Adams_2020} or \(k_{\mathrm{max}} = 0.20h \mathrm{Mpc}^{-1}\) \citep{Koda_2014, Howlett_2017}. We will determine which \(k_{\mathrm{max}}\) to use in Section~\ref{sec:mocks} by choosing the one that returns the mean \(f\sigma_8\) closest to the fiducial value in the mocks. 


The function \(H\) in equation~(\ref{eq:gg_af}) is given by 
\begin{multline}
    H_{p, q}^{l}(\bold{s_1}, \bold{s_2}) = \sum_{l_1, l_2}\frac{4 \pi^2}{(2l_1+1)(2l_2 + 1)}\sum_{m, m_1, m_2} G_{m, m_1, m_2}^{l, l_1, l_2} \\ 
    Y_{l, m}^*(\bold{s_1}-\bold{s_2}) Y_{l_1, m_1}^*(\bold{s_1}) Y_{l_2, m_2}^* (\bold{s_2}) a_{l_1}^{2p} a_{l_2}^{2q},
\label{eq:H}
\end{multline}
where the asterisk denotes the complex conjugate of the spherical harmonics. The Gaunt coefficient \(G\) is given by 
\begin{equation}
\begin{split}
    G^{L_1, L_2, L_3}_{M_1, M_2, M_3} = \sqrt{\frac{(2L_1+1)(2L_2+1)(2L_3+1)}{4 \pi}}\\
    \begin{pmatrix}
    L_1 & L_2 & L_3 \\
    0 & 0 & 0 
    \end{pmatrix}
    \begin{pmatrix}
    L_1 & L_2 & L_3 \\
    M_1 & M_2 & M_3
    \end{pmatrix},
\end{split}
\label{eq:Gaunt}
\end{equation}
where the matrices in equation~(\ref{eq:Gaunt}) represent the Wigner $3j$ symbols. The spherical harmonics functions are denoted by \(Y_{L, M}\) and \(a_l\) is the coefficient of the multipole decomposition of the line-of-sight angle 
\begin{equation}
    a_{l}^{2n} = \frac{2l + 1}{2}\int_{-1}^{1} \mu^{2n}L_l(\mu)d\mu.
    \label{eq:multipole}
\end{equation}

\begin{figure}
	\includegraphics[width=\columnwidth]{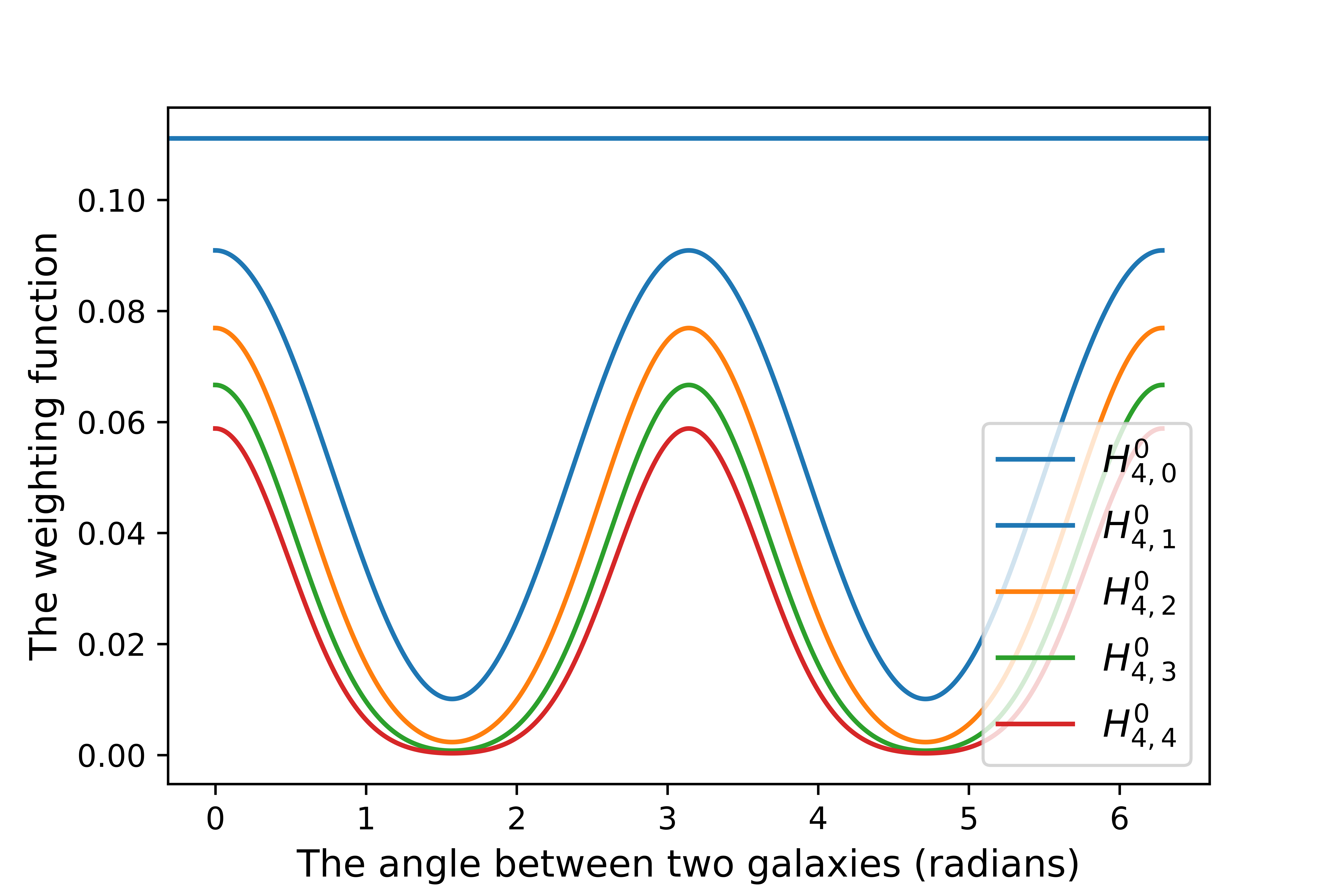}
    \caption{This plot shows how the weighting function \(H\) changes with respect to the angle between two galaxies for \(l = 0\) (monopole and \(l_1, l_2\) are being summed over) and \(p = 4\) (the fourth order term in the Taylor expansion for \(\mu_1\). We varies the value of \(q\) which denotes the \(q^{th}\) order Taylor expansion term for \(\mu_2\). Except when \(q = 0\), the weighting function reaches its maximum when the line-of-sights of two galaxies are parallel or anti-parallel to each other and the minimum is reached when the line-of-sights of two galaxies are perpendicular to each other.}
    \label{fig:Weighting}
\end{figure}

In practice, we evaluate the function equation~(\ref{eq:H}) with \textsc{Mathematica} and sum over different values of $l_{1}$, $l_{2}$. For a given $p$ and $q$, this results in a finite number of nonzero terms as mentioned previously. The final expressions for equation~(\ref{eq:H}) are linear combinations of sine and cosine functions in terms of  \(\theta\) and \(\phi\). Fig~\ref{fig:Weighting} shows how some of the function \(H\) changes with respect to the angle between the two galaxies \(\theta\). You can think about \(H\) as a weighting function here that assigns weights based on the RSD effect. Here, we choose only the monopole terms because they do not depend on \(\phi\). For \(q = 0\), the weighting function is independent of \(\theta\). This is expected since when $p$ or $q$ is zero, we are considering the isotropic component of the clustering, and there is no RSD effect. For higher orders, the weighting function \(H\) reaches the maximum when the line-of-sights to galaxies are either parallel or anti-parallel to each other. It reaches the minimum when the line-of-sights to two galaxies are perpendicular. This is also expected because the finger of god effect elongates the galaxy distribution along the line-of-sight of the observer, while the transverse direction is not affected. It is important to note that at fixed order in $p$ and $q$, the contributions to $\mat{C}_{gg}$ from the velocity-dependent terms enter at one order higher. Hence there is always a $\theta$ dependent contribution to the model that is larger when the galaxies are aligned along the line-of-sight --- for $p=0,q=0$ this is simply the Kaiser effect.

\begin{figure}
	\includegraphics[width=\columnwidth]{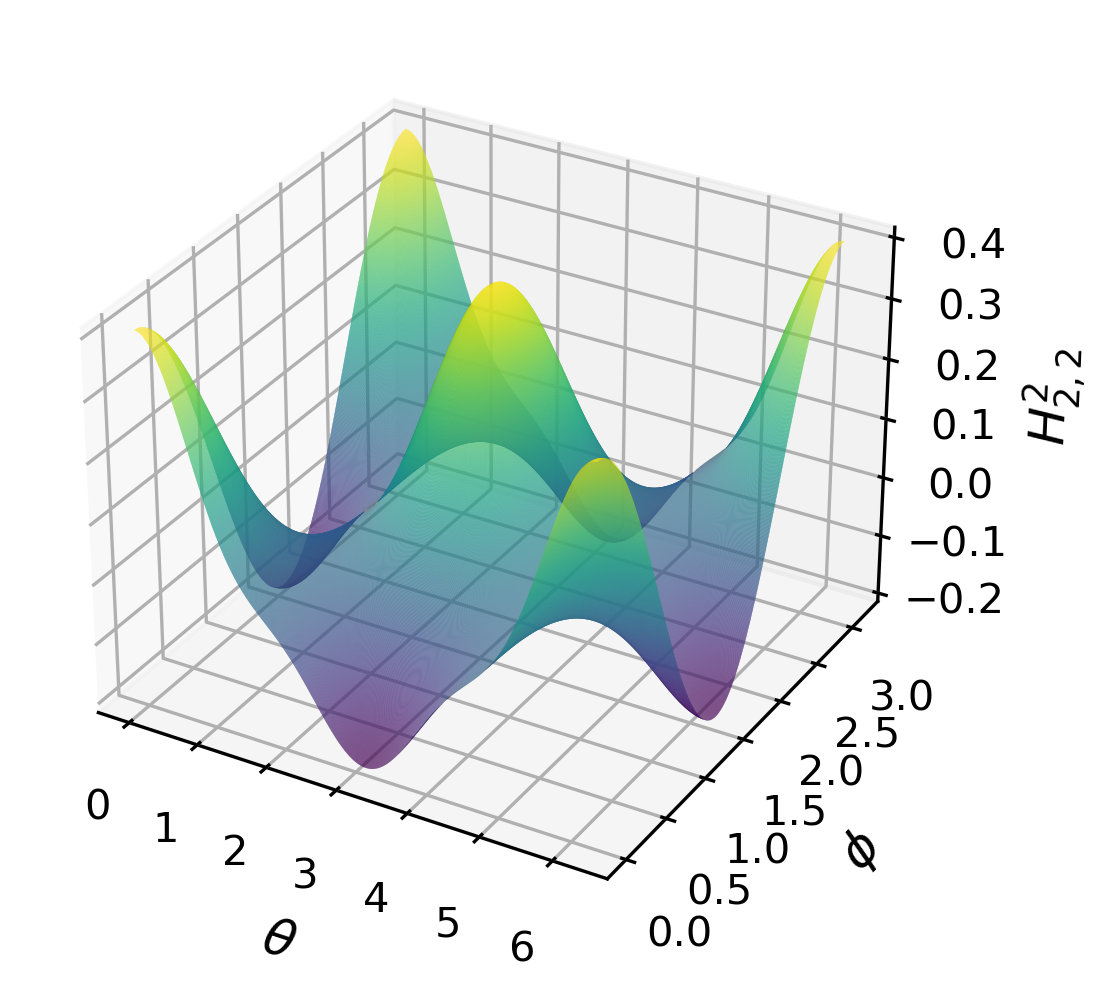}
    \caption{This plot shows how the weighting function changes with respect to \(\theta\) and \(\phi\) for \(l = 2\) (quadrupole and \(l_1, l_2\) are being summed over) and \(p = q = 2\). Similar to the case of the monopole, the weighting function reaches the maximum when the lines-of-sight to two galaxies are parallel or anti-parallel. }
    \label{fig:Weighting_3d}
\end{figure}

The weighting function for other multipoles depends on both \(\theta\) and \(\phi\). Fig~\ref{fig:Weighting_3d} shows the surface plot of the weighting function for the quadrupole (\(l = 2\)) with \(p=q=2\). It shows similar behaviour to Fig~\ref{fig:Weighting}; when two galaxies are parallel (\(\theta = 0, \phi = 0\)) or anti-parallel (\(\theta = \pi, \phi = \frac{\pi}{2}\)), then the weighting function reaches a maximum. The same is true for higher-order multipoles or other values of $p$ and $q$, in that the weighting function still reaches the maximum when the line-of-sights of two galaxies are parallel or anti-parallel, maximising the contribution from RSD. Other than that however, the weighting functions are a complicated function of \(\phi\) and \(\theta\).  

Using a similar approach to the galaxy auto-covariance matrix, the velocity auto-covariance matrix is given by 
\begin{equation}
    \mat{C}_{vv} (s, \sigma_u) = (aHf)^2 \sum_{l} i^{l+2} \xi_{\theta \theta, l}^{-0.5, -0.5, 2}(s, \sigma_u) H_{0.5, 0.5}^{l}(\bold{s_1}, \bold{s_2}). 
    \label{eq:vv_af}
\end{equation}
Although the form of velocity auto-covariance matrix looks at first different from previous derivations \citep{Ma_2011, Adams_2017, Castorina_2020}, we can show they are all mathematically equivalent (see Appendix~\ref{sec:equal} for more detail). This is because all previous derivations of the velocity auto-covariance matrix already have considered the wide-angle effect. Equation~(\ref{eq:vv_be}) shows the highest combined exponent for \(\mu_1\) and \(\mu_2\) is 2, so only \(l = 0, 2\) are non-zero. The multipole order \(l\) here does not depend on the order of the Taylor expansion because equation~(\ref{eq:vv_be}) shows the velocity auto-covariance matrix does not depend on the \(D_g\) damping term. 

Finally, the galaxy-velocity cross-covariance matrix is 
\begin{align}
        & \mat{C}_{gv} (s, \sigma_u) = (aHf) \sum_{p} \frac{(-1)^p}{2^p p!} \sigma_g^{2p} \sum_{l} i^{l+1} \biggl(\xi_{m\theta, l}^{p, -0.5, 1}(s, \sigma_u) \notag \\
        & H_{p, 0.5}^{l}(\bold{s_1}, \bold{s_2}) + f \xi_{\theta \theta, l}^{p, -0.5, 1} (s, \sigma_u)H_{p+1, 0.5}^{l}(\bold{s_1}, \bold{s_2})\biggl)
    \label{eq:gv_af}
\end{align}
and the velocity-galaxy cross-covariance matrix is the transpose of the galaxy-velocity cross-covariance matrix.\footnote{We can deduce this by imposing that the full covariance matrix needs to be symmetric, but Appendix~\ref{sec:appendix_A} analytically shows the cross-covariance matrices are related to each other via a transpose.} Similar to the galaxy auto-covariance matrix, the multipole order \(l\) depends on the order of the Taylor expansion \(p\). The highest order of the multipole is given by \(2(p+1)\). This concludes the formal definition of our model for the covariance matrix for a set of observed galaxy overdensities and peculiar velocities. 

\section{Applying theory to data}
Before we can compute the theoretical covariance matrix for our data, and maximize the likelihood to find the best fitting growth rate, there are a number of other considerations we must make to 1) ensure the model and data are compared on the same footing, and 2) make the comparison computationally feasible. We focus on these two considerations in this section.
\subsection{Modifications to the covariance matrix}
\subsubsection{Log-distance ratio correction}
\citet{Springob_2014} and \citet{Johnson_2014} point out that the uncertainty of the peculiar velocity does not follow a Gaussian distribution. Instead, it follows the log-normal distribution. To solve this issue, they suggested measuring the peculiar velocity with the log-distance ratio \(\eta\) \citep{Springob_2014, Johnson_2014}
\begin{equation}
    \eta = \log_{10}\biggl(\frac{D_z}{D_H}\biggl). 
    \label{eq:log-dist}
\end{equation}
Here, \(D_z\) is the comoving distance deduced from the redshift measurement and \(D_H\) is the true comoving distance from redshift-independent distance measurements. The distribution of log-distance ratio in the SDSS PV catalogue is shown in Fig.~\ref{fig:log_dist_ratio} and it roughly follows the Gaussian distribution. We hence have to convert the model for the velocity auto- and cross-covariance to log-distance ratio. The conversion between the two introduces a conversion factor \citep{Johnson_2014, Watkins_2015}
\begin{equation}
    \kappa(z_{\mathrm{obs}}) = \frac{1}{\ln{10}} \frac{1 + z_{\mathrm{obs}}}{D_z(z_{\mathrm{obs}}) H(z_{\mathrm{obs}})}.
    \label{eq:convert}
\end{equation}
The covariance matrix after the log-distance ratio conversion is then given by 
\begin{equation}
    \mat{C}_{\eta \eta}(\bold{s_1}, \bold{s_2}, \sigma_u) = \kappa(z_{\bold{s_1}})\kappa(z_{\bold{s_2}}) \mat{C}_{vv}(\bold{s_1}, \bold{s_2}, \sigma_u),
    \label{eq:vv_eta}
\end{equation}

\begin{equation}
    \mat{C}_{\eta g}(\bold{s_1}, \bold{s_2}, \sigma_u) = \kappa(z_{\bold{s_1}}) \mat{C}_{vg}(\bold{s_1}, \bold{s_2}, \sigma_u),
    \label{eq:vg_eta}
\end{equation}
and
\begin{equation}
    \mat{C}_{g \eta}(\bold{s_1}, \bold{s_2}, \sigma_u) = \kappa(z_{\bold{s_2}}) \mat{C}_{gv}(\bold{s_1}, \bold{s_2}, \sigma_u).
    \label{eq:gv_eta}
\end{equation}
The galaxy auto-covariance matrix is not affected by this correction because it does not depend on the peculiar velocity data.

\subsubsection{Gridding correction}
\label{sec:gridding}
The SDSS PV catalogue contains more than 34,000 galaxies, making it computationally expensive to compute and fit using the covariance matrix for individual galaxies. Therefore, we choose to reduce the dimensionality of the matrix by gridding the data to reduce the computation time. The centre of each grid cell is then treated as $\boldsymbol{s}_{i}$ in the calculation of the covariance matrix. Additionally, gridding also helps us smooth out the non-linearities in the data, which are typically non-Gaussian, such that it is more appropriate to use the Gaussian likelihood function. 

After gridding the data, the galaxy overdensity of each grid cell is given by 
\begin{equation}
    \delta_g = \frac{N-N_{\mathrm{exp}}\frac{N_{\mathrm{mock}}}{N_{\mathrm{random}}}}{N_{\mathrm{exp}} \frac{N_{\mathrm{mock}}}{N_{\mathrm{random}}}},
    \label{eq:galaxy_density}
\end{equation}
where \(N\) is the number of galaxies in the grid cell and \(N_{\mathrm{exp}}\) is the expected number of galaxies in the same grid cell but with the random catalogue. The number of expected galaxies in the mock \(N_{\mathrm{mock}} = \sum_{\mathrm{grid}} N\) is normalised by the number of galaxies in the random catalogue \(N_{\mathrm{random}} = \sum_{\mathrm{grid}} N_{\mathrm{exp}}\) because each mock may contain a different number of galaxies.  Similarly, the log-distance ratio of each grid cell is the mean of all galaxies' log-distance ratios in the grid cell.

Gridding greatly suppresses the nonlinear power, so we have to modify the covariance matrix models accordingly. This is done by replacing
\begin{equation}
   P_{XY}(k) \rightarrow P_{XY}^{\mathrm{grid}}(k) = P_{XY}(k) \Gamma(k)^2. 
\end{equation}
when computing equation~(\ref{eq:xi}). $\Gamma(k)$ is the angle-averaged, Fourier transform of the gridding kernel with cell edge-length $L$ \citep{Howlett_2017}
\begin{equation}
     \Gamma(k) = \frac{1}{4 \pi} \int_0^{\pi} \int_0^{2 \pi} d\phi d\theta \frac{\sin{k_x} \sin{k_y} \sin{k_z}}{k_x k_y k_z}, 
     \label{eq:grid_window}
\end{equation}     
where the components of the wavevector are given by 
\begin{equation}
    k_x = \frac{k L}{2}\sin{\theta} \cos{\phi}; \quad k_y = \frac{k L}{2}\sin{\theta} \sin{\phi}; \quad k_z = \frac{k L}{2}\cos{\phi}. 
    \label{eq:k_x}
\end{equation} 
The gridding window function is close to one when the wavenumber \(k\) is small and close to zero when \(k\) is larger than the inverse of the length of the grid cell. Gridding also assumes the densities and peculiar velocities of galaxies are continuous. For the galaxy density, this is corrected by the shot noise (see Section~\ref{sec:shotnoise}). For the gridded version of the velocity auto-covariance matrix (\(C_{\eta \eta}^{\rm grid}\)), \citet{Abate_2008} suggest updating the diagonal elements with 
\begin{multline}
    \mat{C}_{\eta \eta}^{\mathrm{grid}}(\bold{s_1}, \bold{s_2}, \sigma_u) \rightarrow \mat{C}_{\eta \eta}^{\mathrm{grid}}(\bold{s_1}, \bold{s_2}, \sigma_u) + \\ \frac{\mat{C}_{\eta \eta}(\bold{s_1}, \bold{s_2}, \sigma_u) - \mat{C}_{\eta \eta}^{\mathrm{grid}}(\bold{s_1}, \bold{s_2}, \sigma_u)}{N}\delta_{D}(\boldsymbol{s}_{1}-\boldsymbol{s}_{2}),
    \label{eq:velocity_sn}
\end{multline}
where \(\mat{C}_{\eta \eta}\) is the non-gridded version of the covariance matrix. The correction is not necessary for the off-diagonal component because they are negligible on small-scales \citep{Abate_2008}. 

To further reduce the dimensionality of the covariance matrix, we delete grid cells of the velocity covariance matrices where there is no galaxies in the catalogue. We will also delete the grid cells of the galaxy covariance matrices where there is no galaxies in the random catalogue. Both of these instances contribute no information to the likelihood function. We use \(30 h^{-1} \mathrm{Mpc}\) grid size when fitting the mocks without applying the Taylor expansion to the logarithmic likelihood function (see \ref{sec:Taylor}). This is because a smaller grid size will take too much time (we estimate it will take more than two weeks) to fit a single mock. After applying the Taylor expansion to the logarithmic likelihood function, the size of the grid cell is \(20 h^{-1} \mathrm{Mpc}\). 

\subsubsection{Shot noise of galaxies and peculiar velocity uncertainties}
\label{sec:shotnoise}
The derivation of the auto-covariance matrices assumes galaxy positions and velocities are continuous fields, whereas they are actually discrete objects. The error introduced is called the shot noise \(\sigma_{\delta_g}\) and can be modelled via Poisson statistics,
\begin{equation}
    \sigma_{\delta_g}(x) = \frac{1}{\sqrt{N_{\mathrm{exp}}(x)}}. 
    \label{eq:shot-noise}
\end{equation}
The shot noise is included in the galaxy auto-covariance matrix by adding it to the diagonal terms 
\begin{equation}
    \mat{C}_{gg}^{\mathrm{err}}(\bold{s_1}, \bold{s_2}, \sigma_g) = \mat{C}_{gg}^{\mathrm{grid}}(\bold{s_1}, \bold{s_2}, \sigma_g) + \sigma^{2}_{\delta_{g}}\delta_{D}(\boldsymbol{s}_{1}-\boldsymbol{s}_{2}).
    \label{eq:gg_sn}
\end{equation}

The error of the log-distance ratio in a given grid cell is treated similarly
\begin{equation}
    \mat{C}_{\eta \eta}^{\mathrm{err}}(\bold{s_1}, \bold{s_2}, \sigma_u) = \mat{C}_{\eta \eta}^{\mathrm{grid}}(\bold{s_1}, \bold{s_2}, \sigma_u) + \sigma^{2}_{\eta}\delta_{D}(\boldsymbol{s}_{1}-\boldsymbol{s}_{2})
\end{equation}
and $\sigma_{\eta}^{\mathrm{grid}}$ contains two independent contributions to the uncertainty of the peculiar velocity $\sigma^{2}_{\eta} = (\sigma_{\eta}^{\mathrm{grid}})^{2} + \kappa(z_{\bold{s_1}})\kappa(z_{\bold{s_2}})\sigma_v^{2}$. The first term arises from the measurement uncertainty of the log-distance ratio of each galaxy in the grid cell and is given as the standard error on the mean
\begin{equation}
    \sigma_{\eta}^{\mathrm{grid}} = \frac{1}{N} \sqrt{\sum_{i} \sigma_{\eta,i}^2}.
    \label{eq:err_grid}
\end{equation}
where each galaxy in the cell has its own uncertainty $\sigma_{\eta,i}$. The second term accounts for the velocity dispersion of galaxies on nonlinear scales. The \(\sigma_v\) parameter is treated as a free parameter in our model.

\subsubsection{Integration bounds}
\citet{Adams_2017} found that there is a significant contribution to the galaxy auto-covariance matrix beyond \(k_{\mathrm{max}} = 0.15h\,\mathrm{Mpc^{-1}}\). They suggested to add an additional integral from \(k_{\mathrm{max}}\) to \(1.0 h\,\mathrm{Mpc^{-1}}\) for the galaxy auto-covariance matrix. This additional integral only acts as a nuisance parameter to increase the value of the galaxy-galaxy auto-covariance matrix, so it does not require complex modelling of the nonlinear power spectrum or the inclusion of redshift space distortions \citep{Adams_2020}. The additional contribution to the covariance matrix is hence given by 
\begin{equation}
    C_{gg}^{\mathrm{add}} =  \sum_{p q} \frac{(-1)^{p+q}}{2^{p+q} p! q!} \sigma_g^{2(p+q)}
        \sum_{l} i^l b_{\mathrm{add}}^2 \xi_{mm, l}^{p, q, 0}(r, 0) H_{p, q}^{l}, 
        \label{eq:b_add}
\end{equation}
where the integration bound for \(\xi\) is from \(k_{\mathrm{max}}\) to \(1.0h\,\mathrm{Mpc^{-1}}\) and \(b_{\mathrm{add}}\) is a free parameter called the additional galaxy bias. Here, \(b_{\mathrm{add}}\) provides an indication of the galaxy bias parameter in the nonlinear scale. 


\subsubsection{Summary}
This subsection has covered a number of modifications made to the model covariance matrices to enable their comparison to the data. To summarise, in fitting the SDSS PV data, we hence replace the covariance matrix used in the likelihood of equation~(\ref{eq:Gaussian_like}) with
\begin{equation}
\boldsymbol{\mathsf{C}} = 
\begin{pmatrix}
    \mat{C}_{gg} & \mat{C}_{gv} \\
    \mat{C}_{vg} & \mat{C}_{vv}
\end{pmatrix}
\rightarrow
\begin{pmatrix}
    \mat{C}_{gg}^{\mathrm{err}}+\mat{C}_{gg}^{\mathrm{add}} & \mat{C}_{g\eta}^{\mathrm{grid}} \\
    \mat{C}_{\eta g}^{\mathrm{grid}} & \mat{C}_{\eta \eta}^{\mathrm{err}}
\end{pmatrix}
.
\end{equation}
The data vector is similarly replaced with
\begin{equation}
\boldsymbol{S} = 
\begin{pmatrix}
    \boldsymbol{\delta}_{g} \\
    \boldsymbol{v}
\end{pmatrix}
\rightarrow
\begin{pmatrix}
    \boldsymbol{\delta}_{g}^{\mathrm{grid}} \\
    \boldsymbol{\eta}^{\mathrm{grid}}
\end{pmatrix}
.
\end{equation}
\subsection{Modifications to the likelihood}

\subsubsection{Zero-point correction}
\label{sec:zeropoint}
The standard method of fitting the fundamental plane requires us to assume that the net velocity inside the survey is zero, which is usually not the case. This `zero-point' then require fixing by comparison to other distance indicators in the distance ladder. In the case of the SDSS PV survey, this was done by cross-matching overlapping galaxies/groups to the Cosmicflows-III catalogue \citep{Tully_2016}. We refer the readers to section 5.4 of \citet{Howlett_2022} for more information on how exactly this was done, but the important point for this work is that such a calibration comes with uncertainty arising from both the small number of objects in common between the two datasets and their individual statistical uncertainties. There is also the chance for hidden systematics to influence either or both of the catalogues. \citet{Howlett_2022} found the zero-point correction for the SDSS PV catalogue with multiple fundamental fit is given by \(y = -0.0037 \pm 0.0040\). We will adopt this measurement in our analysis. 

To account for this uncertainty in our constraints on the growth rate, \citet{Johnson_2014} showed how to marginalize the likelihood function over the uncertainty of zero-point for peculiar velocity surveys. However, to the best of the authors' knowledge, there is no analytical formula for the marginalized likelihood function over the zero-point uncertainty when you combine both the galaxy density and peculiar velocity data. Therefore, we will derive one here. Following from \citet{Johnson_2014}, we assume the zero-point \(y\) has a Gaussian prior
\begin{equation}
    P(y|\sigma_y) = \frac{1}{\sqrt{2\pi} \sigma_y}e^{\frac{-y^2}{2 \sigma_y^2}},
    \label{eq:sigma_y}
\end{equation}
where \(\sigma_y\) here denotes the uncertainty of the zero-point correction. To analytically marginalize over the zero-point uncertainty, we have
\begin{align}
    P(\boldsymbol{S}|m) &= \int dy P(\boldsymbol{S}|m, y) P(y|\sigma_y) \nonumber \\
    &= \int dy \frac{1}{\sqrt{(2\pi)^{n+1} |\mat{C(m)}|} \sigma_y} e^{-\frac{1}{2} (\boldsymbol{S}'^T \mat{C(m)}^{-1}\boldsymbol{S}' + \frac{y^2}{2\sigma_y^2})}.
    \label{eq:MLF}
\end{align}
Due to the zero-point correction, the data vector \(\boldsymbol{S}\) has become \(\boldsymbol{S}' = \boldsymbol{S} + \boldsymbol{x}y\) where \(\boldsymbol{x}\) is zero for the galaxy overdensity data and one for the peculiar velocity data. This is because the zero-point correction will only affect the peculiar velocity data. 


Solving this using properties of Gaussian integrals (see e.g., Appendix A of \citealt{Bridle_2002}) we find,
\begin{equation}
    P(\boldsymbol{S}|m) = \frac{1}{\sqrt{(2\pi)^{n} |\mat{C(m)}|} N_{x}\sigma_{y}}e^{-\frac{1}{2}\left(\boldsymbol{S}^T \mat{C(m)}^{-1} \boldsymbol{S} -\frac{N_y^2}{N_x^2} \right)}
    \label{eq:MLF3}
\end{equation}
where \(N_x = \sqrt{\boldsymbol{x}^T \mat{C(m)}^{-1} \boldsymbol{x} + \frac{1}{\sigma_y^2}}\) and \(N_y = \boldsymbol{S}^T \mat{C(m)}^{-1} \boldsymbol{x}\). Equation~(\ref{eq:MLF3}) can be reduced to the marginalized likelihood function in \citet{Johnson_2014} when there is no galaxy density data. In this case, the vector \(\boldsymbol{x}\) will become a vector of ones. We use this likelihood in place of equation~(\ref{eq:Gaussian_like}) in our fitting, with $\sigma_{y}= 0.0040$.

\subsection{Taylor expansion of the logarithmic likelihood functions}
\label{sec:Taylor}
Although gridding the data as discussed in Section~\ref{sec:gridding} significantly reduces the computational time, it still takes more than one day on a single core to fit a single mock with the exact likelihood function of equation~(\ref{eq:MLF3}). Most of the computational time is spent on calculating the inverse and determinant of the covariance matrix. Both operations scale as \(N_{\rm mat}^3\) where \(N_{\rm mat}\) is the dimension of the covariance matrix. To solve this problem, we first calculate the exact likelihood at the maximum likelihood and then use the Taylor expansion of the likelihood function to interpolate the value of the likelihood during the MCMC sampling. The maximum likelihood is obtained with the optimization, before running a full MCMC chain. 

Our approach will have minimal impact on the uncertainty because the uncertainty of the best-fit is determined by the curvature of the likelihood function around the best-fit. The first derivative of the likelihood function at the maximum likelihood is zero. For a Gaussian likelihood function, the curvature is mainly determined by its second derivative. Therefore, by Taylor expanding the likelihood function up to the second order around the maximum likelihood, we should be able to obtain a similar constraint as with the exact likelihood function. The advantage is that the size of matrices and vectors in the Taylor expansion is given by the number of free parameters which is much less than the size of the covariance matrix. Therefore, it is much faster to evaluate the Taylor expansion of the likelihood than the exact likelihood. The downside of the Taylor expansion method is we need to use optimization to find the maximum likelihood first. We also need to use the exact likelihood during the optimization. However, the optimization requires far fewer iterations than the MCMC, so there is still a huge speed gain with the Taylor expansion method.

To calculate the Taylor expansion of the logarithmic likelihood function, we need to first determine the derivative of the logarithmic likelihood function with respect to each free parameter. From \citet{Petersen_2008}, using the chain rule, we can express the derivative of the logarithmic likelihood function with respect to the free parameters as 
\begin{equation}
    \frac{d\log P(\boldsymbol{S}|\boldsymbol{m})}{d\boldsymbol{m}} = Tr\left(\frac{d\log P(\boldsymbol{S}|\boldsymbol{m})}{d\mat{C(m)}} \frac{d\mat{C(m)}}{d\boldsymbol{m}}\right).
\end{equation}
At the maximum likelihood, we expect the first derivative of the logarithmic likelihood with respect to all the parameters to be zero. However, because we are finding the maximum likelihood numerically, there will be some small numerical errors. Therefore, the first derivative may not necessarily be zero and we have to account for it in the Taylor expansion. We can decompose the marginalized logarithmic likelihood function equation~(\ref{eq:MLF3}) into two parts, one resembling the Gaussian likelihood, and the second arising from our marginalisation over the zero-point
\begin{align}
    \log P(\boldsymbol{S}|\boldsymbol{m}) &= \log P_{\mathrm{G}}(\boldsymbol{S}|\boldsymbol{m}) + \log P_{\mathrm{ZP}}(\boldsymbol{S}|\boldsymbol{m}) \notag \\
    &= -\frac{1}{2} \biggl[n\ln{(2 \pi)} + \ln{(\mat{C(m)})} + \boldsymbol{S}^T \mat{C(m)}^{-1}\boldsymbol{S}\biggl] + \notag \\
    & -\frac{1}{2} \biggl[\ln{(N_x^2 \sigma_y^2)} - \frac{N_y^2}{N_x^2}\biggl].
\end{align}
The first and second derivatives of the logarithmic Gaussian distribution (the first term within the square bracket) with respect to the covariance matrix $\mat{C(m)}$ are well known and so will not be repeated here (see e.g., \citealt{Tegmark_1997}). The derivative of the covariance matrix itself with respect to the parameters is actually also very simple to determine if one realises that the covariance matrix can be decomposed into multiple parts each of which gets multiplied by the parameters at linear or quadratic order (see Appendix~\ref{sec:derivative}). Therefore, we only have to find the first and second derivative for the zero-point correction part of the logarithmic likelihood function \(\log P_{\mathrm{ZP}}(\boldsymbol{S}|\boldsymbol{m})\). Although the calculation is possible analytically, these are somewhat lengthy expressions and so the full expressions are given in Appendix~\ref{sec:second_derivative}.

With the first and second derivatives of the likelihood function in hand, we can then approximate the likelihood at any point in the parameter space $\boldsymbol{m}^{i}$ as
\begin{align}
    & \log P(\boldsymbol{S}|\boldsymbol{m}) = \log P(\boldsymbol{S}|\boldsymbol{m}_{\mathrm{full}}) + \frac{d\log P(\boldsymbol{S}|\boldsymbol{m})}{d\boldsymbol{m}}\biggl|_{\boldsymbol{m}_{\mathrm{full}}} \boldsymbol{\alpha} \nonumber \\
    & +\frac{1}{2} \boldsymbol{\alpha}^T \frac{d^2 \log P(\boldsymbol{S}|\boldsymbol{m})}{d\boldsymbol{m}^2}\biggl|_{\boldsymbol{m}_{\mathrm{full}}} \boldsymbol{\alpha},
\end{align}
where the subscript `full' denotes the nearest point at which we have evaluated the full covariance matrix and likelihood, and \( \boldsymbol{\alpha} = \boldsymbol{m}_{\mathrm{full}} - \boldsymbol{m}\) is the separation between the point and the nearest `full' calculation. After marginalizing over the zero-point, we do not just calculate the Taylor expansion about the maximum likelihood because after marginalizing over the zero-point, the exact likelihood function is no longer Gaussian, so the Taylor expansion is less accurate (see section \ref{sec:after_Marg}). Therefore, we evaluate the Taylor expansion at different points in parameter space after marginalizing over the zero-point. In this case, we pre-compute the first and second derivatives about a suitable number of points (such as the best-fit) before starting MCMC. Therefore, during the MCMC fitting, we only need to calculate \( \boldsymbol{\alpha}\) and its product with the first and second derivatives. However, the Taylor expansion is only accurate for small $\boldsymbol{\alpha}$ after marginalizing over the zero-point, so we will use the mocks to test whether the Taylor expansion of the logarithmic likelihood function is able to return the same posterior as the posterior from using the exact likelihood function.

\section{Testing on SDSS mocks}
\label{sec:mocks}

In this section, we demonstrate the robustness of our theoretical model and fitting methodology on the SDSS PV mock catalogues, which have a known cosmological model. We also use these fits to quantify fiducial values for any parameters that are not varied in our fitting to the data, our systematic error budget, and the expected statistical error for our data.

\subsection{Free parameters}
In total, there are six different free parameters in our model: \(f\sigma_8, \sigma_v, b\sigma_8, b_{\mathrm{add}}\sigma_8, \sigma_u\) and \(\sigma_g\). We neglect the scale-dependent \(\alpha_b\) parameter from \citet{Adams_2020} because we use the same galaxy sample for the galaxy and velocity auto-covariance matrix such that all parameters are determined at the same effective redshift. Of these, only \(\sigma_u\) cannot be varied by simply rescaling a pre-computed version of the appropriate part of the full covariance matrix model with the fiducial cosmological parameters (see Appendix~\ref{sec:derivative}). Therefore, we need to re-evaluate the integral in equation~(\ref{eq:xi}) for different values of \(\sigma_u\). To reduce the computational time, we fix \(\sigma_u\) and justify the fiducial value based on our fits to the mocks. The change in \(f\sigma_8\) due to the change of \(\sigma_u\) will be included as the systematic uncertainty.

We assign flat priors for the remaining five free parameters similar to those assigned by previous works \citep{Johnson_2014, Howlett_2017_b, Adams_2017, Adams_2020}. The prior for the normalized linear growth rate is \(0 \leq f\sigma_8 \leq 1\), the prior for the normalized galaxy bias is \(0 \leq b\sigma_8 \leq 3\), the prior for the nonlinear velocity dispersion is \(0 \mathrm{km s^{-1}}\) \(\leq \sigma_v \leq 5000 \mathrm{km s^{-1}}\), the prior for the additional galaxy bias is \(0 \leq b_{\mathrm{add}}\sigma_8 \leq 10\) and the prior for the finger-of-god damping term is \(0h^{-1} \mathrm{Mpc}\) \(\leq \sigma_g \leq\) \(10h^{-1} \mathrm{Mpc}\).

\subsection{Testing the Taylor expansion method}
\subsubsection{Before applying the zero-point correction.}
One of our first goals is to determine whether using the Taylor expansion of the likelihood function is able to give us the same posterior distribution as using the exact likelihood function. To test this, we fit 5 mocks with a \(30h^{-1} \mathrm{Mpc}\) grid size without marginalizing over zero-point uncertainty, and fixing \(\sigma_u = 13 h^{-1}\mathrm{Mpc}\), \(\sigma_g = 3 h^{-1}\mathrm{Mpc}\) and \(k_{\rm max} = 0.20h^{-1} \mathrm{Mpc}\). Even in this limited case with one less free parameter, the Taylor expansion method gives a substantial speed-up, taking around five minutes to find the posterior distribution on a single core compared to 1.5 days using the full calculation method. Even after taking into account that the Taylor expansion method needs about forty-five minutes to find the best-fit parameters through optimizations and compute the first and second derivative, it is still more than 30 times faster than the full calculation.  

\begin{figure}
	\includegraphics[width=\columnwidth]{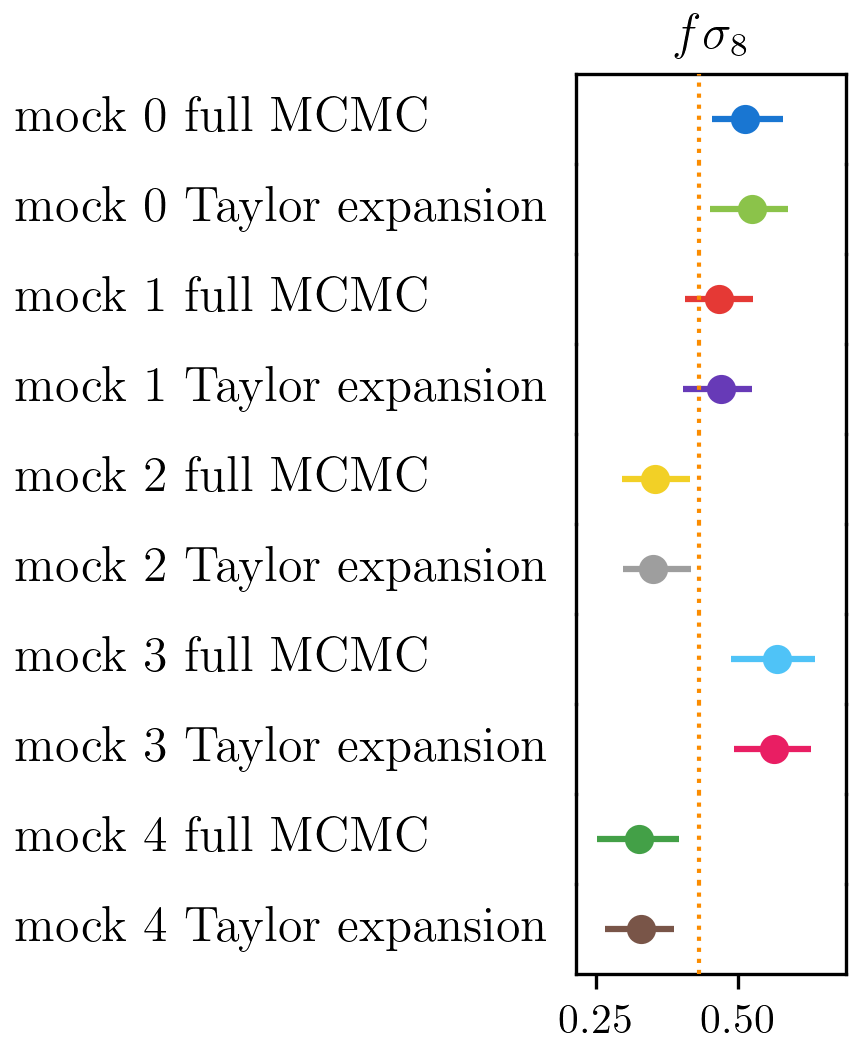}
    \caption{This plot shows the constraints on \(f\sigma_8\) of 5 different mocks with both the exact and Taylor expansion of the likelihood function. It shows the posteriors of \(f\sigma_8\) recovered using the exact likelihood function are the same as the posteriors we get from using the Taylor expansion of the likelihood function. The yellow dashed line shows the fiducial value of \(f\sigma_8 = 0.432\) which is calculated based on the input fiducial cosmological parameters to generate these mocks. }
    \label{fig:approximation}
\end{figure}

Fig~\ref{fig:approximation} shows the recovered marginalised posterior on $f\sigma_{8}$ for these fits. The Taylor expansion method is able to recover almost identical marginalized posteriors as the full MCMC process. The results are also generally in agreement with the expected value for the mock cosmology, but we will explore this further in the next subsection.

\subsubsection{After applying the zero-point correction}
\label{sec:after_Marg}
\begin{figure*}
\begin{multicols}{2}
    \includegraphics[width=\linewidth]{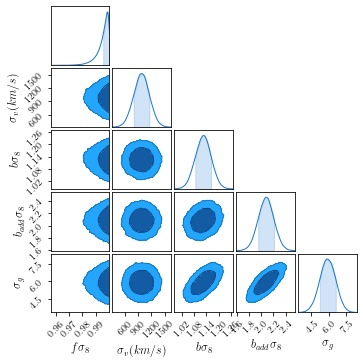}\par 
    \includegraphics[width=\linewidth]{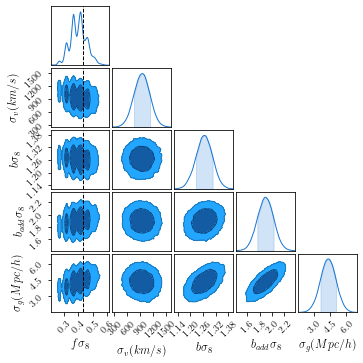}\par
\end{multicols}
\centering
\includegraphics[width=0.5\linewidth]{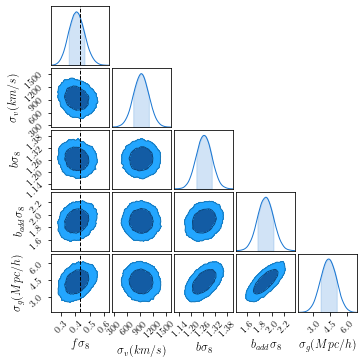}\par
\caption{This figure shows the posteriors of the same mock when calculating the fiducial likelihood at different numbers of points before the MCMC sampling. \emph{Left panel}: we only evaluate Taylor expansion once at the maximum likelihood. The posterior peaks at \(f\sigma_8 = 1\) which is inconsistent with the best-fit from the optimization. This is because the error of the Taylor expansion at \(f\sigma_8 = 1\) is around 10\% which causes it to be larger than the maximum likelihood. Middle panel: we calculate the Taylor expansions at 21 different points with equal intervals in \(f\sigma_8\) space while keeping the other parameters fixed to the best-fit from optimization. The posterior distribution now centers around the fiducial value but it is not smooth. This is because the Taylor expansion is only accurate on the scale smaller than the interval we are using. \emph{Right panel}: we further increase the number of points to 51 and now it is able to return a smooth posterior distribution. For all the mocks we tested, we find 51 points are enough to produce a smooth posterior.}
\label{fig:expansion_error}
\end{figure*}

As a second test of the Taylor expansion methodology, we look at the same mocks but include the effect of marginalizing over the uncertainty of the zero-point correction on the constraint of \(f\sigma_8\). In this case, we found some of the mocks return very high values \(f\sigma_8\). An example is shown in the left panel of Fig~\ref{fig:expansion_error}. Checking the best-fit parameters returned by the optimization algorithm, we find them inconsistent with the maximum likelihood returned by the MCMC. Additionally, the best-fits returned by optimization are consistent with the best-fits without applying the zero-point correction. We hence attribute this problem to numerical inaccuracies in the Taylor expansion far from the best-fit, causing fake maxima in the likelihood function. For example, we compared the exact and Taylor expanded likelihood function for the mock that we used to generate Fig~\ref{fig:expansion_error}, and found good agreement around the maximum likelihood, but an error of $\sim10$\% around \(f\sigma_8 = 1\) causing an overestimation of the log-likelihood here. We also found that except \(f\sigma_8\), the best-fits of other parameters returned by MCMC are consistent with their respective best-fits from optimization and the best-fits without the zero-point correction. This is consistent with our expectation because \(\sigma_g\), \(b\sigma_8\) and \(b_{\mathrm{add}}\sigma_8\) do not depend on the log-distance ratio so the zero-point correction should not impact them. 

We resolve this problem by evaluating the Taylor expansion about multiple different \(f\sigma_8\) values distributed across our prior, and during the MCMC we center the Taylor expansion about the one closest to the proposed \(f\sigma_8\) value. At first, we tried 21 different points with equal spacing. The result is shown in the middle panel of Fig~\ref{fig:expansion_error}, where we now see the posterior is centered around the fiducial value, but the posterior remains noisy because the Taylor expansion is only valid in very small intervals about these points. We then try 51 equally spaced points instead and the result is shown in the right panel of Fig~\ref{fig:expansion_error}. Now the posterior is smooth and centered around the true value as expected. For all the mocks we tested, we find 51 points are good enough to return a smooth unbiased posterior.

\begin{table*}
\centering 
\begin{tabular}{lllllll}
\\
\hline
$\sigma_u (h^{-1} \mathrm{Mpc})$ & $k_{\rm max} (h \mathrm{Mpc}^{-1})$ & mean $f\sigma_8$ & median $f\sigma_8$ & mean uncertainty & median uncertainty & standard deviation \\
\hline
19     & 0.20 & 0.413      & 0.417         & 0.060 (14.6\%) & 0.060 (14.3\%)   & 0.067 (16.2\%)   \\
20     & 0.20 & 0.422       & 0.423         & 0.062 (14.6\%) & 0.061 (14.5\%)   & 0.068 (16.1\%)   \\
21     & 0.20 & 0.429       & 0.429         & 0.063 (14.7\%) & 0.062 (14.6\%)   & 0.068 (15.9\%)   \\
22     & 0.20 & 0.438       & 0.434         & 0.065 (14.8\%) & 0.064 (14.8\%)   & 0.071 (16.3\%)   \\
23     & 0.20 & 0.447       & 0.444         & 0.066 (14.8\%) & 0.066 (14.8\%)   & 0.073 (16.4\%) 
\\
\hline
\end{tabular}
\caption{We tested 201 different mocks with different \(\sigma_u\) and fixed \(k_{\rm max} = 0.20 h \mathrm{Mpc}^{-1}\). The percentage inside the bracket is the relative uncertainty with respect to the mean/median. The relative uncertainty for the standard deviation is calculated with respect to the mean \(f\sigma_8\). Here we show the result for \(\sigma_u\) ranging from \(19h^{-1} \mathrm{Mpc}\) to \(23h^{-1} \mathrm{Mpc}\) because their mean and median \(f\sigma_8\) are the closest ones to the fiducial value (\(f\sigma_8 = 0.432\)). The table demonstrates the mean and median of \(f\sigma_8\) and their uncertainties are very similar. This means there are no significant outliers in the fit. From hereon, we will just use the mean \(f\sigma_8\) and its uncertainty. This table shows \(f\sigma_8\) increases when increasing \(\sigma_u\). This is expected because higher \(\sigma_u\) produces more damping in the power spectrum, so to match the same data, \(f\sigma_8\) has to increase. The table also shows the relative uncertainty of \(f\sigma_8\) does not depend on \(\sigma_u\) as expected. We quantify the effect of the cosmic variance by measuring the standard deviation of the maximum likelihood value of \(f\sigma_8\). We found the cosmic variance has no dependence on \(\sigma_u\) as expected. We found the uncertainty of \(f\sigma_8\) is below its standard deviation which supports \citet{McDonald_2008}'s claim that using multiple tracers of the same underlying matter density field can break the cosmic variance limit.}
\label{tab:sigmau}
\end{table*}

\subsection{Testing the effect of fixing \(\sigma_{u}\)}
We fix \(\sigma_u\) during our analysis, so we want to test how changing \(\sigma_u\) may affect the constraint on \(f\sigma_8\) using the mocks. During this analysis, we fix \(k_{\rm max}\) to \(0.20 h \mathrm{Mpc}^{-1}\). Table~\ref{tab:sigmau} shows the result for \(\sigma_u\) from \(19 h^{-1}\) Mpc to \(23 h^{-1}\) Mpc. We also test other \(\sigma_u\) but these are the ones that produce constraints on \(f\sigma_8\) closest to the fiducial value (\(f\sigma_8 = 0.432\)). Firstly, Table~\ref{tab:sigmau} demonstrates the best-fit \(f\sigma_8\) increases while increasing \(\sigma_u\). This is expected because increasing \(\sigma_u\) means stronger damping in the model power spectrum as shown in equation~(\ref{eq:v_model}). To match the same data, \(f\sigma_8\) has to increase. We calculate both the mean and median of \(f\sigma_8\). Table~\ref{tab:sigmau} shows the mean and median measurement for both \(f\sigma_8\) and its uncertainty are very similar, meaning there are no significant outliers in the data. Therefore, we will calculate the systematic uncertainty introduced using the mean \(f\sigma_8\). The percentage inside the bracket in Table~\ref{tab:sigmau} denotes the relative uncertainty. We find the relative uncertainty is independent of \(\sigma_u\) as expected. We quantify the cosmic variance with the standard deviation of best-fit \(f\sigma_8\) from the mocks \citep{Ruggeri_2020}. The relative uncertainty for the standard deviation is calculated with respect to the mean \(f\sigma_8\). Table ~\ref{tab:sigmau} demonstrates the cosmic variance is independent of \(\sigma_u\) as expected. Additionally, the relative uncertainty of \(f\sigma_8\) is smaller than the relative uncertainty of the cosmic variance. This supports the idea from \citet{McDonald_2008} that using multiple tracers of the same underlying matter density field breaks the cosmic variance limit. Based on Table~\ref{tab:sigmau}, we decide to use \(\sigma_u = 21h^{-1} \mathrm{Mpc}\) to fit the data. The best-fit \(\sigma_u\) we found here is higher than previous results which are usually around \(13h^{-1} \mathrm{Mpc}\) \citep{Koda_2014, Howlett_2017, Adams_2020}. However, \citet{Koda_2014} discovered that the best-fit \(\sigma_u\) depends on the subhalo mass in the simulation. In the SDSS mocks, the subhalo masses are usually larger than the mass ranges used in \citet{Koda_2014}. Therefore, we will expect to find a higher best-fit \(\sigma_u\) with the SDSS mocks.  


%
\subsection{Reduced chi-squared of our model}
To make sure our model is a good fit for the mocks, we also calculated the reduced chi-squared for the 201 mocks with \(\sigma_u = 21h^{-1} \mathrm{Mpc}\) and \(k_{\rm max} = 0.20h^{-1} \mathrm{Mpc}\). The highest reduced chi-squared in the mocks is around 1.19. Different mocks contain different numbers of galaxy overdensities and peculiar velocities measurements, so the degrees of freedom for different mocks are different. However, they are typically around 5500. To understand why some of the mocks return high reduced chi-squared, we fit the galaxy overdensity with the galaxy auto-covariance matrix and the log-distance ratio with the velocity auto-covariance matrix and determine their respective reduced chi-squared. We found the mocks with high reduced chi-squared for the full covariance matrix also had high reduced chi-squared for the galaxy auto-covariance matrix. This indicates our model for the galaxy power spectrum is not a good fit for the galaxy overdensity data. We checked the galaxy overdensity data and found the mocks with high reduced chi-squared also have high galaxy overdensity in some of their grid cells. This means our model is not able to handle the grid cells with high galaxy overdensity. This is expected because the model power spectrum here is developed based on the quasi-linear model, it breaks down when \(\delta_g \gg 1\). We decide to cut out grid cells with \(\delta_g > 20\) in the mocks. We choose twenty because around 20\% of the mocks have no grid cells with \(\delta_g > 20\) and in the mocks with grid cells with \(\delta_g > 20\), we only need to cut out five grid cells on average. The number of grid cells being cut is small compared to the total number of grid cells (around 3,000), so we do not expect this approach to have a huge impact on the statistical uncertainty. 

We refit the mocks and cut out grid cells with \(\delta_g > 20.0\), we find a small change (less than 3\%) in the mean of \(f\sigma_8\) and uncertainty of \(f\sigma_8\). However, the best-fit \(\sigma_u\) becomes 22 \(h^{-1} \rm Mpc\) because its mean (0.430) and median (0.428) of \(f\sigma_8\) are closest to the fiducial value (0.432). The mean uncertainty is 0.062 and the median uncertainty is also 0.062, their respective relative uncertainties are 14.4\% and 14.5\%. The standard deviation is 0.068 or 15.8\% relative to the mean \(f\sigma_8\). Comparing these new values to Table \ref{tab:sigmau}, we conclude the changes are insignificant compared to the mean or median uncertainty. The only advantage of cutting out grid cells with \(\delta_g > 20\) is it improves the reduced chi-square of the fits. Cutting out the high overdensity grid cells has little impact on the degrees of freedom because the number of grid cells that are cut out (usually around 10) is much smaller than the total number of grid cells (around 5500).\footnote{Table \ref{tab:sigmau} demonstrates the fiducial \(f\sigma_8\) is roughly halfway between the mean \(f\sigma_8\) from \(\sigma_u =21 h^{-1} \rm Mpc\) and \(\sigma_u =22 h^{-1} \rm Mpc\). Removing grid cells with high overdensity reduces the overall amplitude of the data power spectrum, this will in terms reduce the value of \(f\sigma_8\), so \(\sigma_u = 22 h^{-1} \rm Mpc\) becomes the best-fit.}

\begin{table*}
\centering
\begin{tabular}{lllllll}
\\
\hline
$\sigma_u (h^{-1} \mathrm{Mpc})$ & $k_{\rm max} (h \mathrm{Mpc}^{-1})$ & mean $f\sigma_8$ & median $f\sigma_8$ & mean uncertainty & median uncertainty & standard deviation \\
\hline
22     & 0.15 & 0.431       & 0.426         & 0.062 (14.5\%) & 0.062 (14.6\%)   & 0.068 (15.8\%)   \\
22     & 0.20 & 0.436       & 0.435         & 0.065 (14.8\%) & 0.064 (14.8\%)   & 0.071 (16.3\%)   \\
22     & 0.25 & 0.425       & 0.425         & 0.064 (15.0\%) & 0.063 (14.9\%)   & 0.071 (16.8\%) 
\\
\hline
\end{tabular}
\caption{This table demonstrates how the constraints on \(f\sigma_8\) change when we change \(k_{\rm max}\). We did not do the overdensity cut for this table because we have shown it has little impact on the mean and uncertainty of \(f\sigma_8\). Similar to Table~\ref{tab:sigmau}, the mean and median of \(f\sigma_8\) agree well with each other, so there is no significant outlier in the data. Additionally, the mean/median relative uncertainty and the standard deviation of \(f\sigma_8\) seem to increase with increasing \(k_{\rm max}\). However, this change is relatively small and has little impact on our final analysis. One may expect increasing \(k_{\rm max}\) will reduce the relative uncertainty because the effective volume of the survey is bigger. However, equation~(\ref{eq:peculiar_velocity}) shows the sensitivity of peculiar velocity to the linear growth rate is inversely proportional to \(k\). Therefore, for smaller scales, the uncertainty of the peculiar velocity dominates so no information was actually added by increasing \(k_{\mathrm{max}}\). Meanwhile, the galaxy power spectrum on the small-scale is dominated by shot noise and the freedom introduced by \(b_{add}\) and \(\sigma_g\) parameters.} 
\label{tab:kmax_change}
\end{table*}

Different to \citet{Adams_2020}, we consider \(k_{\rm max}\) as a model parameter and fix it by considering which \(k_{\rm max}\) returns mean/median \(f\sigma_8\) closest to the fiducial value. Therefore, changing \(k_{\rm max}\) will not introduce additional systematic uncertainty. Table~\ref{tab:kmax_change} demonstrates that increasing \(k_{\rm max}\) does not have a huge impact on the mean and median of \(f\sigma_8\). On the other hand, the relative mean/median uncertainty and the standard deviation increase slightly when increasing \(k_{\rm max}\). However, this change is relatively small and will not have a huge impact on our final analysis. From Table~\ref{tab:kmax_change}, \(k_{\rm max} = 0.15, 0.20, 0.25 h \mathrm{Mpc}^{-1}\) all returns mean/median \(f\sigma_8\) similar to the fiducial value (\(f\sigma_8 = 0.4318\)). We decide to use \(k_{\rm max} = 0.15 h \mathrm{Mpc}^{-1}\) because its mean \(f\sigma_8\) is the closest to the fiducial value and we can compute its covariance matrix faster than the other two \(k_{\rm max}\).

We calculate the Fisher matrix forecast for \(\sigma_u = 22h^{-1} \mathrm{Mpc}\) and \(k_{\rm max} = 0.15 h \mathrm{Mpc}^{-1}\) using the code from \citet{Howlett_2017_b} and find the forecast relative uncertainty of \(f\sigma_8\) is around 12.1\%. The mean relative uncertainty from the mocks is around 14.5\% which is higher than the prediction from the Fisher matrix. However, the Fisher matrix code cannot vary \(\sigma_v\) and \(b_{\rm add}\sigma_8\). Additionally, the code also does not take into account the zero-point correction. If these factors are included, the relative uncertainty from the Fisher matrix forecast will be higher. Therefore, our new method is able to match the prediction from the Fisher matrix.

\begin{figure}
	\includegraphics[width=\columnwidth]{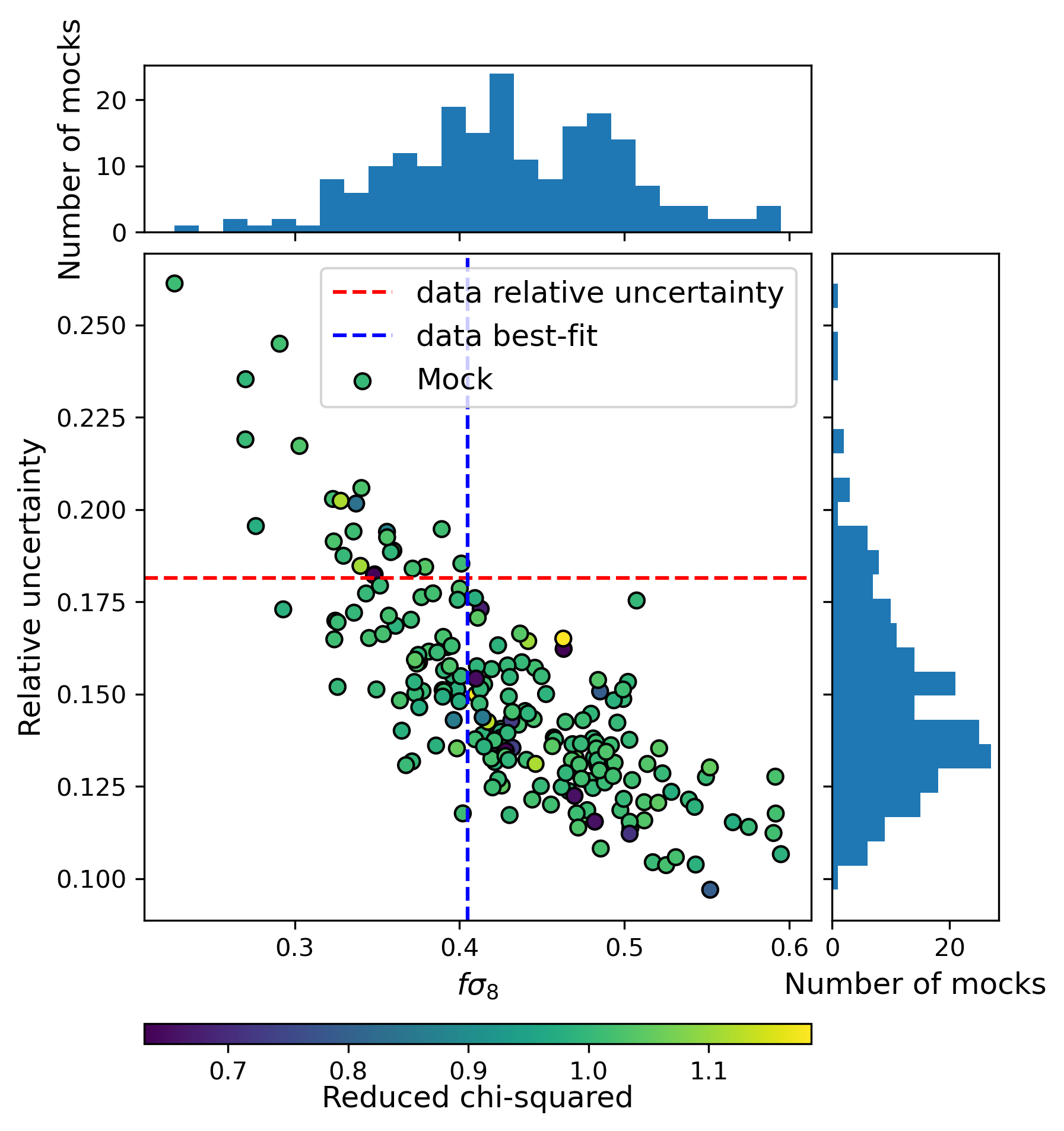}
    \caption{The top panel of this plot shows the histogram of best-fit \(f\sigma_8\) from 201 different mocks with \(k_{\rm max} = 0.15 h \mathrm{Mpc}^{-1}\) and \(\sigma_u = 22 h^{-1} \mathrm{Mpc}\), the bottom panel shows relative uncertainties of best-fit \(f\sigma_8\) from the same 201 different mocks against their respective best-fit \(f\sigma_8\), and the right panel shows the histogram of mocks' relative uncertainties. The reduced chi-squared for the fits with the mocks are shown on the color bar. Most of the fits for the mocks return reduced chi-squared very close to one. The reduced chi-squared for the data fit is 1.03 with the degrees of freedom of 5741. For higher best-fit \(f\sigma_8\), the respective relative uncertainty is lower. This is because \(f\sigma_8\) is similar to a normalization factor for the velocity power spectrum. If the noise in all the mocks is the same, higher \(f\sigma_8\) also indicates a higher signal-to-noise ratio which leads to a lower relative uncertainty. The blue line indicates the best-fit of the SDSS PV data and the red line indicates the relative uncertainty of the SDSS PV data. We can see the relative uncertainty of the data fit is higher than the majority of the mocks, but it is not an outlier.} 
    \label{fig:Relative}
\end{figure}

Table~\ref{tab:sigmau} show the mean and median relative uncertainty with respect to \(\sigma_u\), we also want to determine whether the relative uncertainty depends on the best-fit \(f\sigma_8\) of the mocks. Fig~\ref{fig:Relative} demonstrates that the relative uncertainty reduces when the respective best-fit \(f\sigma_8\) increases. This is because \(f\sigma_8\) acts like a normalization factor to the velocity-divergence power spectrum. A higher \(f\sigma_8\) indicates a higher power spectrum and hence a higher signal-to-noise ratio which consequently reduces the relative uncertainty. Fig~\ref{fig:Relative} also shows for the fits for most of the mocks, the reduced chi-squared is very close to one. 

\section{Data fitting and discussion}
\subsection{Systematic uncertainty}
Besides the statistical uncertainty and the cosmic variance, we also need to consider additional systematic uncertainty from fixing some of our free parameters. When we use the Taylor expansion of the likelihood function, we will fix \(\sigma_u\) to speed up the MCMC fitting. For the MCMC fitting without the Taylor expansion, we also fix \(\sigma_g\) to reduce the computational time. \citet{Adams_2020} demonstrate we can calculate the systematic error introduced by fixing a free parameters \(s\) on the fitting parameter \(\psi\) by 
\begin{equation}
    \sigma_s^2 = \left(\frac{\partial \psi}{\partial s}\right)^2 (\delta s)^2. 
    \label{eq:systematic}
\end{equation}
We can approximate the derivative with the central finite difference method 
\begin{equation}
    \frac{\partial \psi}{\partial s} = \frac{\psi(s+\delta s) - \psi(s-\delta s)}{2 \delta s}. 
    \label{eq:cfdm}
\end{equation}
The total systematic error is then given by 
\begin{equation}
    \sigma_{sys} = \sqrt{\sum_i \sigma_i^2}
    \label{eq:total_sys}
\end{equation}
assuming each systematic uncertainty is independent of each other. The central finite difference method is sensitive to the step size \(\delta s\). A smaller step size will be a better approximation for the derivative, so we fix \(\delta_{\sigma_u} = 1 h^{-1} \mathrm{Mpc}\). Looking at Table \ref{tab:sigmau}, the systematic uncertainty from fixing \(\sigma_u = 22 h^{-1} \rm Mpc\) is 0.009.  


\begin{figure}
	\includegraphics[width=\columnwidth]{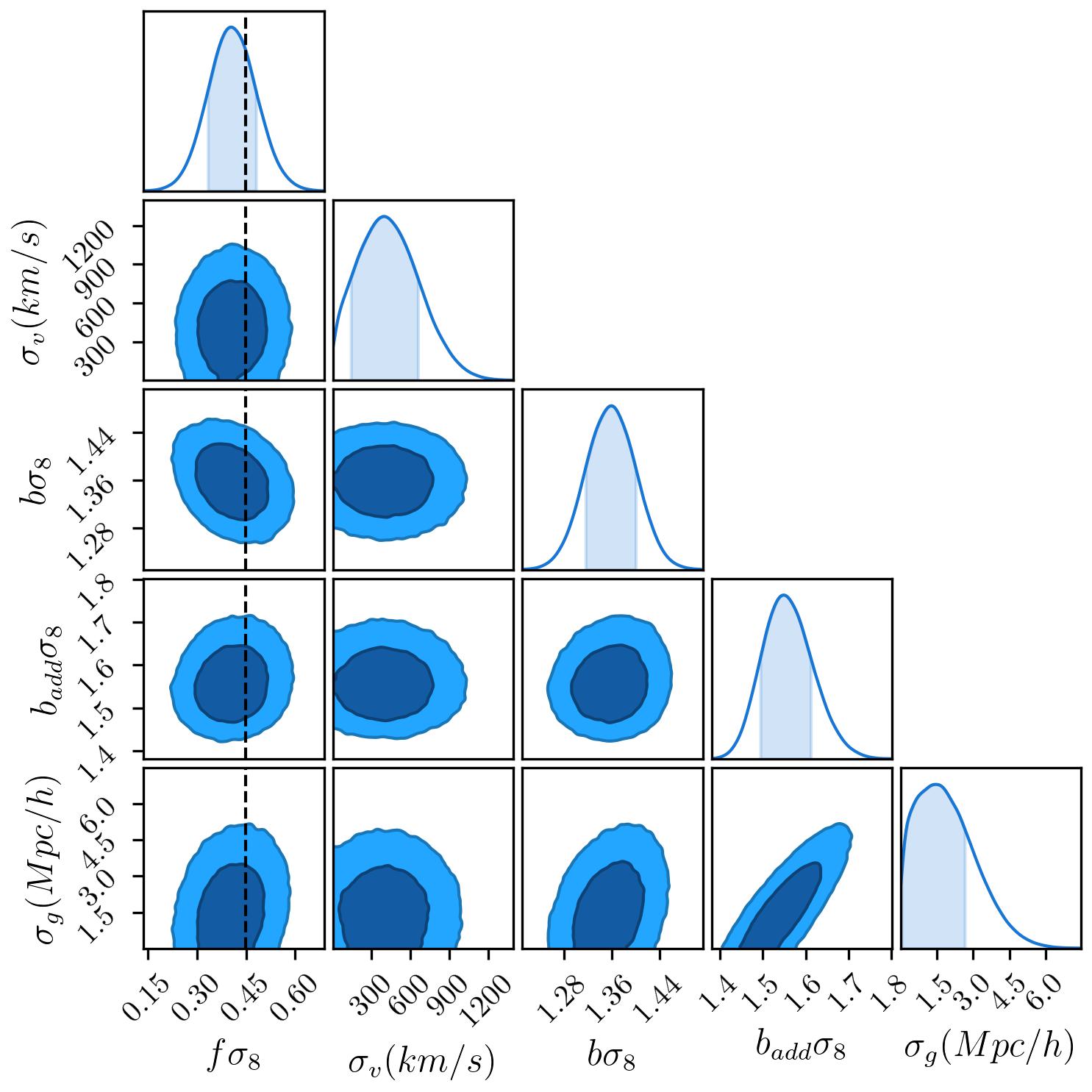}
    \caption{This plot shows our constraints on the free parameters using the SDSS PV data catalogue with \(k_{\rm max} = 0.15 h \mathrm{Mpc}^{-1}\) and \(\sigma_u = 22 h^{-1} \mathrm{Mpc}\) at the effective redshift of 0.073 with overdensity cut at \(\delta_g > 20\). The constraints are \(f\sigma_8 = 0.405_{-0.071}^{+0.076}, b\sigma_8 = 1.359^{+0.041}_{-0.043}, b_{\rm add}\sigma_8 = 1.550_{-0.055}^{+0.062}, \sigma_v = 400_{-260}^{+260}\) km/s and \(\sigma_g = 1.5_{-1.5}^{+1.2} h^{-1} \mathrm{Mpc}\). The prediction from general relativity (dashed line) is calculated from the Planck 2018 cosmology \citep{Planck_2020} at the effective redshift of the data catalogue. The constraint on \(f\sigma_8\) agrees with the prediction from general relativity. Our best-fit \(b_{\rm add}\sigma_8\) is more than 20 standard deviations from zero, so we must include it in the fit in order to return an unbiased result. The nonlinear velocity dispersion \(\sigma_v\) is on the order of a few hundred km/s which is consistent with previous results \citep{Adams_2017, Adams_2020, Howlett_2017}. Additionally, our constraint on \(\sigma_g\) is consistent with the best-fit in \citet{Koda_2014}.} 
    \label{fig:data_fit}
\end{figure}

\subsection{Fitting the SDSS PV data}
The previous section has shown by setting \(\sigma_u = 22 h^{-1} \mathrm{Mpc}\) and \(k_{\rm max} = 0.15 h \mathrm{Mpc}^{-1}\), our model is a good fit for the mocks and we recover unbiased \(f\sigma_8\). Additionally, we also need to remove the grids with galaxy overdensity above 20 in order to obtain a reasonably reduced chi-squared. In total, 12 out of 3126 galaxy overdensity grid cells are removed from the data. Furthermore, different from the mock catalogues which are generated at redshift zero, the effective redshift of the data catalogue is determined to be 0.073. Therefore, we also re-scale the model power spectrum to the effective redshift. Fig.~\ref{fig:data_fit} shows the constraints of all free parameters from the data with the same setting. We find \(f\sigma_8 = 0.405_{-0.071}^{+0.076}\). We calculate the prediction from general relativity (dash line) using the cosmological parameters in Planck 2018 \citep{Planck_2020} at the effective redshift of the data. Our constraint is a bit lower than the prediction from general relativity, but it is still within the error bar. The relative uncertainty is around 18.2\% which is higher than the 14.7\% we find with the mocks as shown in Fig.~\ref{fig:Relative}. This can be explained by Fig.~\ref{fig:Relative} which shows the relative uncertainty increases as the best-fit \(f\sigma_8\) decreases in the mocks. Additionally, we also find the data contain more grid cells that have high overdensity than the average of the mocks which also affects our constraint. This also increases the relative uncertainty of our fit. The constraint on the nonlinear velocity dispersion is \(\sigma_v = 400_{-260}^{+260}\) km/s. This is consistent with previous measurements which are usually around 300 km/s \citep{Howlett_2017, Adams_2017, Adams_2020}. \citet{Howlett_2017} demonstrates gridding the data will significantly loosen the constraint of \(\sigma_v\), so our constraint on \(\sigma_v\) has a large uncertainty. Additionally, the normalized galaxy bias \(b\sigma_8 = 1.441^{+0.043}_{-0.047}\) is also consistent with measurements by \citet{Adams_2017} and \citet{Adams_2020}. Furthermore, we find \(\sigma_g = 1.5_{-1.5}^{+1.2} h^{-1} \mathrm{Mpc}\) which is also consistent with the constraints in \citet{Koda_2014}. Lastly, our constraint on \(b_{\rm add}\sigma_8= 1.550_{-0.055}^{+0.062}\) is more than 20 standard deviations away from zero. This is consistent with \citet{Adams_2020}, so it has to be included in order to return an unbiased constraint on \(f\sigma_8\). The reduced chi-squared of our fit to the data is 1.03 with the degrees of freedom of 5741 which shows our model is a good fit for the data.\footnote{For comparison, the reduced chi-squared for the fit to the data before cutting out the high overdensity grid cells is 1.08 with 5753 degrees of freedom.} 

We fix \(\sigma_u\) during fitting so we have to take this into account by adding in the systematic uncertainty due to fixing \(\sigma_u\). The final constraint on the normalized linear growth rate is \(f\sigma_8 = 0.405_{-0.071}^{+0.076} ( \mathrm{stat}) \pm 0.009 (\mathrm{sys})\). The statistical uncertainty here is larger than the spread of the mean from the mocks. Therefore, the cosmic variance is already included by the statistical uncertainty \citep{Ruggeri_2020}. The total relative uncertainty is around 18.2\%. The previous constraints by \citet{Adams_2020} with the 6dFGSv sample using the maximum likelihood method has relative uncertainty around 20.9\%. \footnote{In \citet{Adams_2020}, they have twice as many galaxy samples (70467) as SDSS, but only a quarter as many velocity samples (8885) as SDSS. In total, they have 79352 data points while we have 68118 data points. Additionally, they did not take the zero-point correction into account, so their error bar is underestimated. Although they included \(k_{\rm max}\) as a free parameter, but Table~\ref{tab:kmax_change} demonstrates \(k_{\rm max}\) has little impact on \(f\sigma_8\).} Similarly, \citet{Said_2020} used the velocity reconstruction technique to constrain \(f\sigma_8\) with data from SDSS and 6dFGSv. They found \(f\sigma_8 = 0.338 \pm 0.027\). Our measurement is consistent with \citet{Said_2020} but our relative uncertainty is two times larger. This is because the velocity reconstruction technique requires far fewer free parameters and it assumes the reconstruction technique is valid on all scales. Additionally, they also use much more redshifts than our work to perform the reconstruction calculation. In comparison, we only fit our model up to \(k_{\rm max} = 0.15 h \mathrm{Mpc}^{-1}\). Consequently, the maximum likelihood method gives a much lower uncertainty than the velocity reconstruction approach. 

\section{Conclusion}
This work improves the previous work by \citet{Adams_2020} by taking the wide-angle effect into account. We achieve this by Taylor expanding the damping function \(D_g\) about the cosine of the line-of-sight angle \(\mu\). The exact solution is an infinite summation but we show just the first few orders of the Taylor expansion will be able to recover the covariance matrices accurately. We choose to include the first four orders of the Taylor expansion and calculate the corresponding covariance matrices. In this work, we speed up the MCMC by first calculating the likelihood at some fiducial points and then using the Taylor expansion of the likelihood function to interpolate the values of the likelihood during the MCMC sampling. We discover this method significantly reduces the computational time compared to using the exact likelihood function. This is because the size of matrices and vectors in the Taylor expansion is given by the number of free parameters which is much smaller than the number of data points. More importantly, it recovers almost an identical posterior distribution for the free parameters as using the exact likelihood function. Through testing the mocks, we found the best-fit \(\sigma_u\) is \(22 h^{-1} \mathrm{Mpc}\). This value is higher than the result in \citet{Koda_2014} from N-body simulation because the masses of sub-halos in our mocks are much higher. We also choose to fit the data with \(k_{\rm max} = 0.15 h \mathrm{Mpc}^{-1}\). Furthermore, we remove the grids with overdensity above 20 because our method fails to model such high galaxy overdensity on small scales. By setting \(\sigma_u = 22 h^{-1} \mathrm{Mpc}\) and \(k_{\rm max} = 0.15 h\mathrm{Mpc}^{-1}\), the reduced chi-squared for the fits with the mocks are mostly close to one, indicating our model is a good fit to the mocks. Using the same setting, we fit the SDSS PV data catalogue and find \(f\sigma_8 = 0.405_{-0.071}^{+0.076}\) (stat) \(\pm 0.009\) (sys). The constraint on \(f\sigma_8\) is slightly lower than the general relativity prediction calculated using the Planck 2018 cosmological parameters \citep{Planck_2020}. Nonetheless, the general relativity prediction is within the error bar range of our best-fit, so our result is consistent with the prediction from general relativity. The relative uncertainty of \(f\sigma_8\) is smaller than \citet{Adams_2020} which used the same method but using the 6dFGSv data. However, our relative uncertainty of \(f\sigma_8\) is about two times larger than the relative uncertainty of \(f\sigma_8\) in \citet{Said_2020} who uses a combination of SDSS PV catalogue up to redshift of 0.055 and the 6dFGSv data. This is because the velocity reconstruction method in \citet{Said_2020} has far fewer free parameters and can be applied to all scales and they also use more galaxy overdensity data to apply the reconstruction technique. In the future, our new method can be applied to future peculiar velocity surveys to obtain a stronger constraint. 

\section*{Acknowledgements}
This research was supported by the Australian Government through the Australian
Research Council’s Laureate Fellowship funding scheme (project FL180100168). YL is the recipient of the Graduate School Scholarship of The University of Queensland. This research has made use of NASA's Astrophysics Data System Bibliographic Services and the \texttt{astro-ph} pre-print archive at \url{https://arxiv.org/}, the {\sc matplotlib} plotting library \citep{Hunter2007}, the {\sc pvista} plotting library \citep{Sullivan_2019}, and the {\sc chainconsumer} and {\sc emcee} packages \citep{Hinton2016, ForemanMackey2013}. The computation is performed at the Getafix supercomputer at the University of Queensland. 

\section*{Data Availability}
The code used in this research is published here \url{https://github.com/YanxiangL/Peculiar_velocity_fitting} and the SDSS PV catalogue data, mock, and random files are on: \url{https://zenodo.org/record/6640513}.



\bibliographystyle{mnras}
\bibliography{example} 

\begin{thebibliography}{}
\makeatletter
\relax
\def\mn@urlcharsother{\let\do\@makeother \do\$\do\&\do\#\do\^\do\_\do\%\do\~}
\def\mn@doi{\begingroup\mn@urlcharsother \@ifnextchar [ {\mn@doi@}
  {\mn@doi@[]}}
\def\mn@doi@[#1]#2{\def\@tempa{#1}\ifx\@tempa\@empty \href
  {http://dx.doi.org/#2} {doi:#2}\else \href {http://dx.doi.org/#2} {#1}\fi
  \endgroup}
\def\mn@eprint#1#2{\mn@eprint@#1:#2::\@nil}
\def\mn@eprint@arXiv#1{\href {http://arxiv.org/abs/#1} {{\tt arXiv:#1}}}
\def\mn@eprint@dblp#1{\href {http://dblp.uni-trier.de/rec/bibtex/#1.xml}
  {dblp:#1}}
\def\mn@eprint@#1:#2:#3:#4\@nil{\def\@tempa {#1}\def\@tempb {#2}\def\@tempc
  {#3}\ifx \@tempc \@empty \let \@tempc \@tempb \let \@tempb \@tempa \fi \ifx
  \@tempb \@empty \def\@tempb {arXiv}\fi \@ifundefined
  {mn@eprint@\@tempb}{\@tempb:\@tempc}{\expandafter \expandafter \csname
  mn@eprint@\@tempb\endcsname \expandafter{\@tempc}}}

\bibitem[\protect\citeauthoryear{{Abate}, {Bridle}, {Teodoro}, {Warren}  \&
  {Hendry}}{{Abate} et~al.}{2008}]{Abate_2008}
{Abate} A.,  {Bridle} S.,  {Teodoro} L. F.~A.,  {Warren} M.~S.,   {Hendry} M.,
  2008, \mn@doi [\mnras] {10.1111/j.1365-2966.2008.13637.x}, \href
  {https://ui.adsabs.harvard.edu/abs/2008MNRAS.389.1739A} {389, 1739}

\bibitem[\protect\citeauthoryear{{Adams} \& {Blake}}{{Adams} \&
  {Blake}}{2017}]{Adams_2017}
{Adams} C.,  {Blake} C.,  2017, \mn@doi [\mnras] {10.1093/mnras/stx1529}, \href
  {https://ui.adsabs.harvard.edu/abs/2017MNRAS.471..839A} {471, 839}

\bibitem[\protect\citeauthoryear{{Adams} \& {Blake}}{{Adams} \&
  {Blake}}{2020}]{Adams_2020}
{Adams} C.,  {Blake} C.,  2020, \mn@doi [\mnras] {10.1093/mnras/staa845}, \href
  {https://ui.adsabs.harvard.edu/abs/2020MNRAS.494.3275A} {494, 3275}

\bibitem[\protect\citeauthoryear{Aghanim et~al.,}{Aghanim
  et~al.}{2020}]{Planck_2020}
Aghanim N.,  et~al., 2020, \mn@doi [Astronomy & Astrophysics]
  {10.1051/0004-6361/201833910}, 641, A6

\bibitem[\protect\citeauthoryear{Blake et~al.,}{Blake
  et~al.}{2013}]{Blake_2013}
Blake C.,  et~al., 2013, \mn@doi [Monthly Notices of the Royal Astronomical
  Society] {10.1093/mnras/stt1791}, 436, 3089–3105

\bibitem[\protect\citeauthoryear{Boruah, Hudson  \& Lavaux}{Boruah
  et~al.}{2020}]{Boruah_2020}
Boruah S.~S.,  Hudson M.~J.,   Lavaux G.,  2020, \mn@doi [Monthly Notices of
  the Royal Astronomical Society] {10.1093/mnras/staa2485}, 498, 2703–2718

\bibitem[\protect\citeauthoryear{{Bridle}, {Crittenden}, {Melchiorri},
  {Hobson}, {Kneissl}  \& {Lasenby}}{{Bridle} et~al.}{2002}]{Bridle_2002}
{Bridle} S.~L.,  {Crittenden} R.,  {Melchiorri} A.,  {Hobson} M.~P.,  {Kneissl}
  R.,   {Lasenby} A.~N.,  2002, \mn@doi [\mnras]
  {10.1046/j.1365-8711.2002.05709.x}, \href
  {https://ui.adsabs.harvard.edu/abs/2002MNRAS.335.1193B} {335, 1193}

\bibitem[\protect\citeauthoryear{{Brout} et~al.,}{{Brout}
  et~al.}{2022}]{Brout_2022}
{Brout} D.,  et~al., 2022, arXiv e-prints, \href
  {https://ui.adsabs.harvard.edu/abs/2022arXiv220204077B} {p. arXiv:2202.04077}

\bibitem[\protect\citeauthoryear{Burkey \& Taylor}{Burkey \&
  Taylor}{2004}]{Burkey_2004}
Burkey D.,  Taylor A.~N.,  2004, \mn@doi [Monthly Notices of the Royal
  Astronomical Society] {10.1111/j.1365-2966.2004.07192.x}, 347, 255–268

\bibitem[\protect\citeauthoryear{Campbell et~al.,}{Campbell
  et~al.}{2014}]{Campbell_2014}
Campbell L.~A.,  et~al., 2014, \mn@doi [Monthly Notices of the Royal
  Astronomical Society] {10.1093/mnras/stu1198}, 443, 1231

\bibitem[\protect\citeauthoryear{Carrick, Turnbull, Lavaux  \& Hudson}{Carrick
  et~al.}{2015}]{Carrick_2015}
Carrick J.,  Turnbull S.~J.,  Lavaux G.,   Hudson M.~J.,  2015, \mn@doi
  [Monthly Notices of the Royal Astronomical Society] {10.1093/mnras/stv547},
  450, 317–332

\bibitem[\protect\citeauthoryear{Castorina \& White}{Castorina \&
  White}{2018}]{Castorina_2018}
Castorina E.,  White M.,  2018, \mn@doi [Monthly Notices of the Royal
  Astronomical Society] {10.1093/mnras/sty410}

\bibitem[\protect\citeauthoryear{Castorina \& White}{Castorina \&
  White}{2020}]{Castorina_2020}
Castorina E.,  White M.,  2020, \mn@doi [Monthly Notices of the Royal
  Astronomical Society] {10.1093/mnras/staa2129}, 499, 893–905

\bibitem[\protect\citeauthoryear{Davis \& Scrimgeour}{Davis \&
  Scrimgeour}{2014}]{Davis_2014}
Davis T.~M.,  Scrimgeour M.~I.,  2014, \mn@doi [Monthly Notices of the Royal
  Astronomical Society] {10.1093/mnras/stu920}, 442, 1117

\bibitem[\protect\citeauthoryear{De~Felice \& Tsujikawa}{De~Felice \&
  Tsujikawa}{2010}]{De_Felice_2010}
De~Felice A.,  Tsujikawa S.,  2010, \mn@doi [Living Reviews in Relativity]
  {10.12942/lrr-2010-3}, 13

\bibitem[\protect\citeauthoryear{{Dekel} \& {Lahav}}{{Dekel} \&
  {Lahav}}{1999}]{Dekel_1999}
{Dekel} A.,  {Lahav} O.,  1999, \mn@doi [\apj] {10.1086/307428}, \href
  {https://ui.adsabs.harvard.edu/abs/1999ApJ...520...24D} {520, 24}

\bibitem[\protect\citeauthoryear{{Djorgovski} \& {Davis}}{{Djorgovski} \&
  {Davis}}{1987}]{Djorgovski_1987}
{Djorgovski} S.,  {Davis} M.,  1987, \mn@doi [\apj] {10.1086/164948}, \href
  {https://ui.adsabs.harvard.edu/abs/1987ApJ...313...59D} {313, 59}

\bibitem[\protect\citeauthoryear{{Dressler}, {Lynden-Bell}, {Burstein},
  {Davies}, {Faber}, {Terlevich}  \& {Wegner}}{{Dressler}
  et~al.}{1987}]{Dressler_1987}
{Dressler} A.,  {Lynden-Bell} D.,  {Burstein} D.,  {Davies} R.~L.,  {Faber}
  S.~M.,  {Terlevich} R.,   {Wegner} G.,  1987, \mn@doi [\apj]
  {10.1086/164947}, \href
  {https://ui.adsabs.harvard.edu/abs/1987ApJ...313...42D} {313, 42}

\bibitem[\protect\citeauthoryear{Dupuy, Courtois  \& Kubik}{Dupuy
  et~al.}{2019}]{Dupuy_2019}
Dupuy A.,  Courtois H.~M.,   Kubik B.,  2019, \mn@doi [Monthly Notices of the
  Royal Astronomical Society] {10.1093/mnras/stz901}, 486, 440–448

\bibitem[\protect\citeauthoryear{Dvali, Gabadadze  \& Porrati}{Dvali
  et~al.}{2000}]{Dvali_2000}
Dvali G.,  Gabadadze G.,   Porrati M.,  2000, \mn@doi [Physics Letters B]
  {10.1016/s0370-2693(00)00669-9}, 485, 208–214

\bibitem[\protect\citeauthoryear{{Foreman-Mackey}, {Hogg}, {Lang}  \&
  {Goodman}}{{Foreman-Mackey} et~al.}{2013}]{ForemanMackey2013}
{Foreman-Mackey} D.,  {Hogg} D.~W.,  {Lang} D.,   {Goodman} J.,  2013, \mn@doi
  [\pasp] {10.1086/670067}, \href
  {http://adsabs.harvard.edu/abs/2013PASP..125..306F} {125, 306}

\bibitem[\protect\citeauthoryear{Gil-Marín et~al.,}{Gil-Marín
  et~al.}{2015}]{Gil_Mar_n_2015}
Gil-Marín H.,  et~al., 2015, \mn@doi [Monthly Notices of the Royal
  Astronomical Society] {10.1093/mnras/stv1359}, 452, 1914–1921

\bibitem[\protect\citeauthoryear{{Hinton}}{{Hinton}}{2016}]{Hinton2016}
{Hinton} S.~R.,  2016, \mn@doi [The Journal of Open Source Software]
  {10.21105/joss.00045}, \href
  {https://ui.adsabs.harvard.edu/abs/2016JOSS....1...45H} {1, 00045}

\bibitem[\protect\citeauthoryear{Howlett}{Howlett}{2019}]{Howlett_2019}
Howlett C.,  2019, \mn@doi [Monthly Notices of the Royal Astronomical Society]
  {10.1093/mnras/stz1403}, 487, 5209–5234

\bibitem[\protect\citeauthoryear{{Howlett}, {Staveley-Smith}  \&
  {Blake}}{{Howlett} et~al.}{2017a}]{Howlett_2017_b}
{Howlett} C.,  {Staveley-Smith} L.,   {Blake} C.,  2017a, \mn@doi [\mnras]
  {10.1093/mnras/stw2466}, \href
  {https://ui.adsabs.harvard.edu/abs/2017MNRAS.464.2517H} {464, 2517}

\bibitem[\protect\citeauthoryear{{Howlett} et~al.,}{{Howlett}
  et~al.}{2017b}]{Howlett_2017}
{Howlett} C.,  et~al., 2017b, \mn@doi [\mnras] {10.1093/mnras/stx1521}, \href
  {https://ui.adsabs.harvard.edu/abs/2017MNRAS.471.3135H} {471, 3135}

\bibitem[\protect\citeauthoryear{Howlett, Said, Lucey, Colless, Qin, Lai, Tully
   \& Davis}{Howlett et~al.}{2022}]{Howlett_2022}
Howlett C.,  Said K.,  Lucey J.~R.,  Colless M.,  Qin F.,  Lai Y.,  Tully
  R.~B.,   Davis T.~M.,  2022, \mn@doi [Monthly Notices of the Royal
  Astronomical Society] {10.1093/mnras/stac1681}, 515, 953

\bibitem[\protect\citeauthoryear{{Hunter}}{{Hunter}}{2007}]{Hunter2007}
{Hunter} J.~D.,  2007, \mn@doi [Computing in Science and Engineering]
  {10.1109/MCSE.2007.55}, \href
  {http://adsabs.harvard.edu/abs/2007CSE.....9...90H} {9, 90}

\bibitem[\protect\citeauthoryear{Huterer, Shafer, Scolnic  \& Schmidt}{Huterer
  et~al.}{2017}]{Huterer_2017}
Huterer D.,  Shafer D.~L.,  Scolnic D.~M.,   Schmidt F.,  2017, \mn@doi
  [Journal of Cosmology and Astroparticle Physics]
  {10.1088/1475-7516/2017/05/015}, 2017, 015–015

\bibitem[\protect\citeauthoryear{Jackson}{Jackson}{1972}]{Jackson_1972}
Jackson J.~C.,  1972, \mn@doi [Monthly Notices of the Royal Astronomical
  Society] {10.1093/mnras/156.1.1p}, 156, 1P

\bibitem[\protect\citeauthoryear{{Johnson} et~al.,}{{Johnson}
  et~al.}{2014}]{Johnson_2014}
{Johnson} A.,  et~al., 2014, \mn@doi [\mnras] {10.1093/mnras/stu1615}, \href
  {https://ui.adsabs.harvard.edu/abs/2014MNRAS.444.3926J} {444, 3926}

\bibitem[\protect\citeauthoryear{{Kaiser}}{{Kaiser}}{1986}]{Kaiser_1986}
{Kaiser} N.,  1986, \mn@doi [\mnras] {10.1093/mnras/222.2.323}, \href
  {https://ui.adsabs.harvard.edu/abs/1986MNRAS.222..323K} {222, 323}

\bibitem[\protect\citeauthoryear{{Koda} et~al.,}{{Koda}
  et~al.}{2014}]{Koda_2014}
{Koda} J.,  et~al., 2014, \mn@doi [\mnras] {10.1093/mnras/stu1610}, \href
  {https://ui.adsabs.harvard.edu/abs/2014MNRAS.445.4267K} {445, 4267}

\bibitem[\protect\citeauthoryear{Lilow \& Nusser}{Lilow \&
  Nusser}{2021}]{lilow_2021}
Lilow R.,  Nusser A.,  2021, Constrained realizations of 2MRS density and
  peculiar velocity fields: growth rate and local flow (\mn@eprint {arXiv}
  {2102.07291})

\bibitem[\protect\citeauthoryear{Linder \& Cahn}{Linder \&
  Cahn}{2007}]{Linder_2007}
Linder E.~V.,  Cahn R.~N.,  2007, \mn@doi [Astroparticle Physics]
  {10.1016/j.astropartphys.2007.09.003}, 28, 481–488

\bibitem[\protect\citeauthoryear{Ma, Gordon  \& Feldman}{Ma
  et~al.}{2011}]{Ma_2011}
Ma Y.-Z.,  Gordon C.,   Feldman H.~A.,  2011, \mn@doi [Physical Review D]
  {10.1103/physrevd.83.103002}, 83

\bibitem[\protect\citeauthoryear{Massey et~al.,}{Massey
  et~al.}{2007}]{Massey_2007}
Massey R.,  et~al., 2007, \mn@doi [The Astrophysical Journal Supplement Series]
  {10.1086/516599}, 172, 239–253

\bibitem[\protect\citeauthoryear{{McDonald} \& {Seljak}}{{McDonald} \&
  {Seljak}}{2009a}]{McDonald_2008}
{McDonald} P.,  {Seljak} U.,  2009a, \mn@doi [\jcap]
  {10.1088/1475-7516/2009/10/007}, \href
  {https://ui.adsabs.harvard.edu/abs/2009JCAP...10..007M} {2009, 007}

\bibitem[\protect\citeauthoryear{McDonald \& Seljak}{McDonald \&
  Seljak}{2009b}]{McDonald_2009}
McDonald P.,  Seljak U.,  2009b, \mn@doi [Journal of Cosmology and
  Astroparticle Physics] {10.1088/1475-7516/2009/10/007}, 2009, 007–007

\bibitem[\protect\citeauthoryear{Nusser}{Nusser}{2017}]{Nusser_2017}
Nusser A.,  2017, \mn@doi [Monthly Notices of the Royal Astronomical Society]
  {10.1093/mnras/stx1225}, 470, 445–454

\bibitem[\protect\citeauthoryear{Park}{Park}{2000}]{Park_2000}
Park C.,  2000, \mn@doi [Monthly Notices of the Royal Astronomical Society]
  {10.1111/j.1365-8711.2000.03886.x}, 319, 573

\bibitem[\protect\citeauthoryear{Park \& Park}{Park \& Park}{2006}]{Park_2006}
Park C.-G.,  Park C.,  2006, \mn@doi [The Astrophysical Journal]
  {10.1086/498258}, 637, 1

\bibitem[\protect\citeauthoryear{{Peacock} \& {Dodds}}{{Peacock} \&
  {Dodds}}{1994}]{Peacock_1994}
{Peacock} J.~A.,  {Dodds} S.~J.,  1994, \mn@doi [\mnras]
  {10.1093/mnras/267.4.1020}, \href
  {https://ui.adsabs.harvard.edu/abs/1994MNRAS.267.1020P} {267, 1020}

\bibitem[\protect\citeauthoryear{Petersen \& Pedersen}{Petersen \&
  Pedersen}{2008}]{Petersen_2008}
Petersen K.~B.,  Pedersen M.~S.,  2008, The Matrix Cookbook, \url
  {http://www2.imm.dtu.dk/pubdb/p.php?3274}

\bibitem[\protect\citeauthoryear{Qin, Howlett  \& Staveley-Smith}{Qin
  et~al.}{2019}]{Qin_2019}
Qin F.,  Howlett C.,   Staveley-Smith L.,  2019, \mn@doi [Monthly Notices of
  the Royal Astronomical Society] {10.1093/mnras/stz1576}, 487, 5235–5247

\bibitem[\protect\citeauthoryear{{Ruggeri} \& {Blake}}{{Ruggeri} \&
  {Blake}}{2020}]{Ruggeri_2020}
{Ruggeri} R.,  {Blake} C.,  2020, \mn@doi [\mnras] {10.1093/mnras/staa2540},
  \href {https://ui.adsabs.harvard.edu/abs/2020MNRAS.498.3744R} {498, 3744}

\bibitem[\protect\citeauthoryear{{Said}, {Colless}, {Magoulas}, {Lucey}  \&
  {Hudson}}{{Said} et~al.}{2020}]{Said_2020}
{Said} K.,  {Colless} M.,  {Magoulas} C.,  {Lucey} J.~R.,   {Hudson} M.~J.,
  2020, \mn@doi [Monthly Notices of the Royal Astronomical Society]
  {10.1093/mnras/staa2032}, \href
  {https://ui.adsabs.harvard.edu/abs/2020MNRAS.tmp.2141S} {}

\bibitem[\protect\citeauthoryear{Shiraishi, Akitsu  \& Okumura}{Shiraishi
  et~al.}{2021}]{Maresuke_2021}
Shiraishi M.,  Akitsu K.,   Okumura T.,  2021, Alcock-Paczynski effects on
  wide-angle galaxy statistics (\mn@eprint {arXiv} {2103.08126})

\bibitem[\protect\citeauthoryear{{Springob} et~al.,}{{Springob}
  et~al.}{2014}]{Springob_2014}
{Springob} C.~M.,  et~al., 2014, \mn@doi [\mnras] {10.1093/mnras/stu1743},
  \href {https://ui.adsabs.harvard.edu/abs/2014MNRAS.445.2677S} {445, 2677}

\bibitem[\protect\citeauthoryear{Strauss \& Willick}{Strauss \&
  Willick}{1995}]{Strauss_1995}
Strauss M.~A.,  Willick J.~A.,  1995, \mn@doi [Physics Reports]
  {10.1016/0370-1573(95)00013-7}, 261, 271–431

\bibitem[\protect\citeauthoryear{{Sullivan} \& {Kaszynski}}{{Sullivan} \&
  {Kaszynski}}{2019}]{Sullivan_2019}
{Sullivan} C.,  {Kaszynski} A.,  2019, \mn@doi [The Journal of Open Source
  Software] {10.21105/joss.01450}, \href
  {https://ui.adsabs.harvard.edu/abs/2019JOSS....4.1450S} {4, 1450}

\bibitem[\protect\citeauthoryear{Taylor \& Watts}{Taylor \&
  Watts}{2001}]{Taylor_2001}
Taylor A.,  Watts P.,  2001, \mn@doi [Monthly Notices of the Royal Astronomical
  Society] {10.1046/j.1365-8711.2001.04874.x}, 328, 1027–1038

\bibitem[\protect\citeauthoryear{{Tegmark}}{{Tegmark}}{1997}]{Tegmark_1997}
{Tegmark} M.,  1997, \mn@doi [\prl] {10.1103/PhysRevLett.79.3806}, \href
  {https://ui.adsabs.harvard.edu/abs/1997PhRvL..79.3806T} {79, 3806}

\bibitem[\protect\citeauthoryear{{Tully} \& {Fisher}}{{Tully} \&
  {Fisher}}{1977}]{Tully_1977}
{Tully} R.~B.,  {Fisher} J.~R.,  1977, \aap, \href
  {https://ui.adsabs.harvard.edu/abs/1977A&A....54..661T} {500, 105}

\bibitem[\protect\citeauthoryear{Tully, Courtois  \& Sorce}{Tully
  et~al.}{2016}]{Tully_2016}
Tully R.~B.,  Courtois H.~M.,   Sorce J.~G.,  2016, \mn@doi [The Astronomical
  Journal] {10.3847/0004-6256/152/2/50}, 152, 50

\bibitem[\protect\citeauthoryear{{Turner}, {Blake}  \& {Ruggeri}}{{Turner}
  et~al.}{2021}]{Turner_2021}
{Turner} R.~J.,  {Blake} C.,   {Ruggeri} R.,  2021, \mn@doi [\mnras]
  {10.1093/mnras/stab212}, \href
  {https://ui.adsabs.harvard.edu/abs/2021MNRAS.502.2087T} {502, 2087}

\bibitem[\protect\citeauthoryear{Watkins \& Feldman}{Watkins \&
  Feldman}{2015}]{Watkins_2015}
Watkins R.,  Feldman H.~A.,  2015, \mn@doi [Monthly Notices of the Royal
  Astronomical Society] {10.1093/mnras/stv651}, 450, 1868

\bibitem[\protect\citeauthoryear{{Zel'Dovich}}{{Zel'Dovich}}{1970}]{Zeldovich_1970}
{Zel'Dovich} Y.~B.,  1970, \aap, \href
  {https://ui.adsabs.harvard.edu/abs/1970A&A.....5...84Z} {500, 13}

\makeatother
\end{thebibliography}




\appendix

\section{The derivation of the full covariance matrix}
\label{sec:appendix_A}
\subsection{The derivation of the galaxy auto-covariance matrix}
From equation~(\ref{eq:gg_be}), the galaxy auto-covariance matrix is given by 
\begin{equation}
\begin{split}
        \mat{C}_{gg} = \int \frac{d^3 k}{(2 \pi)^3} e^{i \bold{k} \cdot \bold{r}} \biggl(b^2 P_{mm} + bf \mu_1^2 P_{m\theta} +  bf \mu_2^2 P_{m\theta} + \\
        f^2 \mu_1^2 \mu_2^2 P_{\theta \theta}\biggl) e^{-\frac{(k \mu_1 \sigma_g)^2}{2}} e^{-\frac{(k\mu_2\sigma_g)^2}{2}}.
\end{split}
\label{eq:gg_be_again}
\end{equation}
The Taylor expansion of \(D_g\) is given by 
\begin{equation}
    e^{-\frac{k^2\sigma_g^2 (\mu_1^2 + \mu_2^2)}{2}} = \sum_{p = 0}^{\infty} \sum_{q = 0}^{\infty} \frac{(-1)^{p+q}}{2^{p+q}p! q!} k^{2(p+q)} \sigma_g^{2(p+q)} \mu_1^{2p} \mu_2^{2q}.
    \label{eq:Taylor_D_g}
\end{equation}
We can decompose the line-of-sight angle \(\mu\) with the multipole expansion in equation~(\ref{eq:multipole}). Additionally, we can decompose \(e^{i \bold{k} \cdot \bold{r}} \) with the plane wave decomposition 
\begin{equation}
    e^{i \bold{k} \cdot \bold{r}} = \sum_l i^l (2l+1) j_l(kr) L_l(\hat{k} \cdot \hat{r})
    \label{eq:pwd}
\end{equation}
The galaxy auto-covariance matrix then becomes 
\begin{align}
\begin{split}
    \mat{C}_{gg} = \sum_{p,q} \frac{(-1)^{p+q}}{2^{p+q}p! q!} \int \frac{d^3 k}{(2 \pi)^3} \sum_l i^l (2l+1) j_l(kr) L_l(\hat{k} \cdot \hat{r})\\
    k^{2(p+q)} \sigma_g^{2(p+q)}
    \sum_{l_1, l_2} b^2 P_{mm} a_{l_1}^{2p} a_{l_2}^{2q} L_{l_1}(\hat{k} \cdot \hat{s_1}) L_{l_2}(\hat{k} \cdot \hat{s_2})\\
    + bf P_{m\theta} L_{l_1}(\hat{k} \cdot \hat{s_1}) L_{l_2}(\hat{k} \cdot \hat{s_2}) (a_{l_1}^{2p+2} a_{l_2}^{2q} + a_{l_1}^{2p} a_{l_2}^{2q+2}) + \\
    f^2 P_{\theta \theta} L_{l_1}(\hat{k} \cdot \hat{s_1}) L_{l_2}(\hat{k} \cdot \hat{s_2}) a_{l_1}^{2p+2} a_{l_2}^{2q+2}.
\end{split}
\label{eq:gg_pwd}
\end{align}

We can further simplify the equation with the spherical harmonics addition theorem 
\begin{equation}
    L_l(\hat{k} \cdot \hat{r}) = \frac{4 \pi}{2l + 1}\sum_{m = -l}^l Y_{lm}(\hat{k}) Y_{lm}(\hat{r})^*. 
    \label{eq:shat}
\end{equation}
Transferring into the spherical coordinate and substituting equation~(\ref{eq:shat}), the galaxy auto-covariance matrix now becomes 
\begin{equation}
\begin{split}
    \mat{C}_{gg} = \sum_{p,q} \frac{(-1)^{p+q}}{2^{p+q}p! q!} \int \frac{k^2 dk}{2 \pi^2} \sum_{l, l_1, l_2} i^l j_l(kr) \sum_{m, m_1, m_2} \\
    \int_{0}^{\pi} \int_{0}^{2 \pi} \biggl[Y_{lm}(\hat{k})
    Y_{l_1 m_1}(\hat{k}) Y_{l_2 m_2}(\hat{k}) \sin{\theta} d\phi d\theta\biggl]\\ 
    Y_{lm}(\hat{s})^* Y_{l_1 m_1}(\hat{s_1})^* Y_{l_2 m_2}(\hat{s_2})^*
    \frac{4\pi^2}{(2l_1+1)(2l_2+1)} \\
    \Bigg(b^2 P_{mm}a_{l_1}^{2p} a_{l_2}^{2q} + bf P_{m\theta}(a_{l_1}^{2p+2} a_{l_2}^{2q} + a_{l_1}^{2p} a_{l_2}^{2q+2})\\
    +f^2 P_{\theta \theta} a_{l_1}^{2p+2} a_{l_2}^{2q+2} \Bigg).
\end{split} 
\label{eq:gg_shat}
\end{equation}
Recalling the definition for the Gaunt coefficient is 
\begin{equation}
    \int_{0}^{\pi} \int_{0}^{2\pi} Y_{LM}(\hat{k}) Y_{L_1 M_1}(\hat{k}) Y_{L_2 M_2}(\hat{k}) \sin{\theta} d\phi d\theta = G_{L, L_1, L_2}^{M, M_1, M_2}.  \label{eq:Gaunt_again}
\end{equation}
Substituting the definition for the Gaunt coefficient, we have 
\begin{equation}
\begin{split}
    \mat{C}_{gg} = \sum_{p,q} \frac{(-1)^{p+q}}{2^{p+q}p! q!} \int \frac{k^2 dk}{2 \pi^2} \sum_{l, l_1, l_2} i^l j_l(kr) \sum_{m, m_1, m_2} \\
    G_{l, l_1, l_2}^{m, m_1, m_2} 
    Y_{lm}(\hat{s})^* Y_{l_1 m_1}(\hat{s_1})^* Y_{l_2 m_2}(\hat{s_2})^*
    \frac{(4\pi)^2}{(2l_1+1)(2l_2+1)} \\
    \Bigg(b^2 P_{mm}a_{l_1}^{2p} a_{l_2}^{2q} + bf P_{m\theta}(a_{l_1}^{2p+2} a_{l_2}^{2q} + a_{l_1}^{2p} a_{l_2}^{2q+2}) \\
    +f^2 P_{\theta \theta} a_{l_1}^{2p+2} a_{l_2}^{2q+2} \Bigg).
\end{split} 
\label{eq:gg_Gaunt}
\end{equation}
After substituting the definition for the \(\xi\) function and the \(H\) function from equation~(\ref{eq:xi}) and equation~(\ref{eq:H}), we recover equation~(\ref{eq:gg_af}). 

\subsection{Cross-covariance matrices}
From equation~(\ref{eq:gv_be}) and equation~(\ref{eq:vg_be}), the covariance matrix for the galaxy-velocity cross-covariance matrix and the velocity galaxy cross-covariance matrix are 
\begin{equation}
    \begin{split}
        \mat{C}_{gv} = - i a H f \int \frac{d^3 k}{(2 \pi)^3} e^{i \bold{k} \cdot \bold{r}} \frac{\mu_2}{k} D_u(k, \sigma_u) \\
        (b P_{m\theta} + f \mu_1^2 P_{\theta \theta})D_g(k, \sigma_g, \mu_1) 
    \end{split}
    \label{eq:gv_be_again}
\end{equation}
and
\begin{equation}
    \begin{split}
        \mat{C}_{vg} = i a H f \int \frac{d^3 k}{(2 \pi)^3} e^{i \bold{k} \cdot \bold{r}} \frac{\mu_1}{k} D_u(k, \sigma_u) \\
        (b P_{m\theta} + f \mu_2^2 P_{\theta \theta})D_g(k, \sigma_g, \mu_2) 
    \end{split}
    \label{eq:vg_be_again}
\end{equation}
respectively. 
Applying the Taylor expansion to \(D_g(k, \sigma_g, \mu_i) \) and the plane wave decomposition to \(e^{i \bold{k} \cdot \bold{r}}\), the galaxy-velocity cross-covariance matrix and the velocity-galaxy cross-covariance matrix become
\begin{equation}
    \begin{split}
         \mat{C}_{gv} = - i a H f \sum_{p} \int \frac{d^3 k}{(2 \pi)^3} \frac{(-1)^p}{2^p p!} D_u(k, \sigma_u)\\
         \sum_l i^l j_l(ks)(2l+1) L_l(\hat{k} \cdot \hat{s}) k^{2p - 1} \sigma_g^{2p} \sum_{l_1, l_2} \\
         \left(bf P_{m\theta}a_{l_1}^{2p} a_{l_2} +f^2 P_{\theta \theta} a_{l_1}^{2p+2} a_{l_2}\right)L_{l_1}(\hat{k} \cdot \hat{s_1}) L_{l_2}(\hat{k} \cdot \hat{s_2}) 
    \end{split}
    \label{eq:gv_pwd}
\end{equation}
and 
\begin{equation}
    \begin{split}
         \mat{C}_{vg} = i a H f \sum_{p} \int \frac{d^3 k}{(2 \pi)^3} \frac{(-1)^p}{2^p p!} D_u(k, \sigma_u) \\
         \sum_l i^l j_l(ks)(2l+1) L_l(\hat{k} \cdot \hat{s}) k^{2p - 1} \sigma_g^{2p} \sum_{l_1, l_2} \\
         \left(bf P_{m\theta}a_{l_1} a_{l_2}^{2p+2} +f^2 P_{\theta \theta} a_{l_1} a_{l_2}^{2p+2}\right)L_{l_1}(\hat{k} \cdot \hat{s_1}) L_{l_2}(\hat{k} \cdot \hat{s_2}) 
    \end{split}
    \label{eq:vg_pwd}
\end{equation}
respectively. If we swapped \(\hat{s_1}\) and \(\hat{s_2}\) which is equivalent to taking the transpose for the galaxy-velocity auto-covariance matrix, we have \(\hat{s} \rightarrow -\hat{s}\) and the \(a_{l_1}\) and \(a_{l_2}\) are swapped. For the cross-covariance matrices, only the odd multipoles will survive because the required values of \(l\) are determined by the total power of the product of \(\mu_1\) and \(\mu_2\) in the anisotropic power spectra and they are odd numbers \citep{Adams_2020}. The odd orders of Legendre polynomials are odd functions, which cancels out the minus sign in equation~(\ref{eq:gv_pwd}). Therefore, when we take the transpose of the galaxy-velocity cross-covariance matrix, we recover the velocity-galaxy cross-covariance matrix. From hereon, we will only show the derivation for the galaxy-velocity cross-covariance matrix. 

Applying the spherical harmonics addition theorem and the definition of the Gaunt coefficient, the galaxy-velocity auto-covariance matrix is given by 
\begin{equation}
    \begin{split}
         \mat{C}_{gv} = - i a H f \sum_{p} \int \frac{k^2 dk}{4 \pi} \frac{(-1)^p}{2^p p!} D_u(k, \sigma_u) k^{2p - 1} \sigma_g^{2p} \\
         \sum_{l, l_1, l_2} i^l j_l(ks)
         \sum_{m, m_1, m_2} G_{l, l_1, l_2}^{m, m_1, m_2} Y{lm}(\hat{s})^* Y_{l_1, m_1}(\hat{s_1})^* Y_{l_2, m_2}(\hat{s_2})^* \\
         \frac{(4\pi)^2}{(2l_1+1)(2l_2+1)}
         \left(bf P_{m\theta}a_{l_1}^{2p} a_{l_2} +f^2 P_{\theta \theta} a_{l_1}^{2p+2} a_{l_2}\right). 
    \end{split}
    \label{eq:gv_Gaunt}
\end{equation}
Substituting in equation~(\ref{eq:xi}) and equation~(\ref{eq:H}), we recover equation~(\ref{eq:gv_af}). Similarly, we can show the covariance matrix for the velocity-galaxy cross-covariance matrix is 
\begin{equation}
    \begin{split}
        \mat{C}_{vg} (s, \sigma_u) = -(aHf) \sum_{p} \frac{i (-1)^p}{2^p p!} \sigma_g^{2p} \sum_{l} i^l \\
        \Bigg(b \xi_{m\theta, l}^{p, -0.5, 1}(s, \sigma_u)
        H_{0.5, p}^{l} + f \xi_{\theta \theta, l}^{p, -0.5, 1} H_{0.5, p+1}^{l}\Bigg). 
    \end{split}
\label{eq:vg_af_again}
\end{equation}
Similar to the galaxy auto-covariance matrix, we only include the first four terms of the Taylor expansion (\(p <= 3\)). This means we have to sum up to \(l = 7\). 
\subsection{Velocity auto-covariance matrix}
From equation~(\ref{eq:vv_be}), the velocity auto-covariance matrix is 
\begin{equation}
    \begin{split}
        \mat{C}_{vv} = \int \frac{d^3 k}{(2 \pi)^3} e^{i \bold{k} \cdot \bold{r}} \frac{(aHf)^2}{k^2} D_u^2 (k, \sigma_u) \mu_1 \mu_2.
    \end{split}
    \label{eq:vv_be_again}
\end{equation}
Applying the multipole expansion and the plane wave decomposition, the velocity auto-covariance matrix is given by 
\begin{equation}
    \begin{split}
        \mat{C}_{vv} =  (aHf)^2\int \frac{d^3 k}{(2\pi)^3} \frac{D_u^2}{k^2} \sum_l i^l (2l+1) j_l(ks) L_l(\hat{k} \cdot \hat{s}) \\ 
        \sum_{l_1, l_2} a_{l_1} a_{l_2} L_{l_1}(\hat{k} \cdot \hat{s_1}) L_{l_2}(\hat{k} \cdot \hat{s_2}). 
    \end{split}
    \label{eq:vv_pwd}
\end{equation}
Applying the spherical harmonics addition theorem and the definition of the Gaunt integral, the velocity auto-covariance matrix becomes 
\begin{equation}
    \begin{split}
        \mat{C}_{vv} =  (aHf)^2\int \frac{d^3 k}{(2\pi)^3} \frac{D_u^2}{k^2} \sum_{l, l_1, l_2} i^l j_l(ks) \sum_{m, m_1, m_2} G_{l, l_1, l_2}^{m, m_1, m_2} \\
        Y_{lm}(\hat{s})^* Y_{l_1 m_1}(\hat{s_1})^* Y_{l_2, m_2} (\hat{s_2}) a_{l_1} a_{l_2}^*. 
    \end{split}
    \label{eq:vv_shat}
\end{equation}
Substituting equation~(\ref{eq:xi}) and equation~(\ref{eq:H}), we can find the velocity auto-covariance matrix is given by equation~(\ref{eq:vv_af}). There is no \(D_g\) terms in the velocity auto-covariance matrix, so we don't have to apply the Taylor expansion. The highest order of the product of \(\mu_1\) and \(\mu_2\) is two, so the velocity auto-covariance matrix only depends on the monopole and the quadrupole. 

\section{Comparison with previous results}
\label{sec:appendix_B}
\citet{Adams_2017} derives the formulae for the covariance matrices that neglect the effect from RSD. This is equivalent to setting \(\sigma_u = \sigma_g = 0\) for our covariance matrices and assuming the galaxy overdensity is given by 
\begin{equation}
    \delta_g = b\delta_m.
    \label{eq:galaxy_density_2017}
\end{equation}
This section will show that by setting such limits, our equations reduce to the equations in \citet{Adams_2017}. 

\subsection{Galaxy auto-covariance matrix} 
The galaxy auto-covariance matrix in \citet{Adams_2017} is 
\begin{equation}
    \mat{C}_{gg} = \frac{b^2}{2\pi^2} \int dk P_{mm}(k) k^2 j_0(kr). 
\end{equation}
With \(\sigma_g = 0\), only terms with \(p = q = 0\) will survive. Using the definition of galaxy overdensity in equation~(\ref{eq:galaxy_density_2017}), the terms with the cross-power spectrum and the velocity divergence auto-power spectrum will vanish. Additionally, since  \(p = q = 0\), the highest order of \(\mu\) is zero. Therefore, we only have to include the \(l = 0\) term. This gives 
\begin{equation}
    \mat{C}_{gg} = b^2 \xi_{mm, 0}^{0, 0, 0} H_{0, 0}^{0}. 
    \label{eq:gg_simp_2017}
\end{equation}
From \textsc{Mathematica}, \(H_{0, 0}^{0} = 1\). If we substitute in the definition for \(\xi\) function, we find 
\begin{equation}
    \mat{C}_{gg} = \frac{b^2}{2\pi^2} \int dk P_{mm}(k) k^2 j_0(kr),
    \label{eq:gg_2017}
\end{equation}
which is consistent with the formula from \citet{Adams_2017}. 

\subsection{Galaxy-velocity cross-covariance matrix}
From \citet{Adams_2017}, the galaxy-velocity cross-covariance matrix without the RSD correction is 
\begin{equation}
    \mat{C}_{gv} = -\frac{aHfb}{2\pi^2} \int P_{m \theta} k (\hat{s_2} \cdot \hat{r}) j_1(kr). 
    \label{eq:gv_2017}
\end{equation}

Similar to the galaxy auto-covariance matrix, only terms with \(p = q = 0\) and \(l = 1\) survive. With the definition of the galaxy overdensity in equation~(\ref{eq:galaxy_density_2017}), we only have the cross-power spectrum. After simplification, the galaxy-velocity auto-covariance matrix is 
\begin{equation}
    \mat{C}_{gv} = aHbf \xi_{m \theta, 1}^{0, -0.5, 1} H_{0, 0.5}^{1}. 
    \label{eq:gv_simp}
\end{equation}
From \textsc{Mathematica}, \(H_{0, 0.5}^{1} = -\cos{(\phi + \frac{\theta}{2})}\). From Fig~\ref{fig:Geometry}, \(\psi = \pi - (\phi + \frac{\theta}{2})\) and using the fact that \(\cos{(\pi - \theta)} = -\cos{\theta}\), we have 
\begin{align}
    \mat{C}_{gv} &= -aHbf \xi_{m \theta, 1}^{0, -0.5, 1} \cos{\psi} \nonumber \\
    &= -aHbf \xi_{m \theta, 1}^{0, -0.5, 1} (\hat{s_2} \cdot \hat{r}) 
    \label{eq:gv_simp_2017}
\end{align}
because the dot product of two unit vectors gives the cosine of the angle between them. Substituting in \(\xi_{m \theta, 1}^{0, -0.5, 1} = \int \frac{k dk}{2 \pi^2} P_{m \theta} j_1(kr)\), we recover equation~(\ref{eq:gv_2017}). 

\subsection{Velocity auto-covariance matrix}
\label{sec:equal}
The velocity auto-covariance matrix does not depend on the \(D_g\) terms, so we expect our equation is mathematically equivalent to the velocity auto-covariance matrix formula in \citet{Adams_2017} 
\begin{align}
        & \mat{C}_{vv} = \frac{(aHf)^2}{2\pi^2} \int dk P_{\theta \theta} \nonumber \\ 
        & \left(\frac{1}{3}\cos{\theta}[j_0(kr)-2j_2(kr)] + \frac{s_1 s_2}{r^2} j_2(kr) (\sin{\theta})^2\right).   
\label{eq:vv_2017}
\end{align}
After simplifying equation~(\ref{eq:vv_af}) with \textsc{Mathematica}, our derivation gives 
\begin{align}
         & \mat{C}_{vv} = \frac{(aHf)^2}{2\pi^2} \int dk P_{\theta \theta} \nonumber \\
         & \left(\frac{1}{3}\cos{\theta}j_0(kr) - \frac{3\cos{2\phi} + \cos{\theta}}{6} j_2(kr)\right).
    \label{eq:vv_full}
\end{align}
Thus we only need to show 
\begin{equation}
    - \frac{3\cos{2\phi} + \cos{\theta}}{6} = \frac{s_1 s_2}{r^2} (\sin{\theta})^2-\frac{2}{3}\cos{\theta}.
    \label{eq:vv_equal}
\end{equation}
Starting from the left-hand side of equation~(\ref{eq:vv_equal}), by applying the angular bisector theorem and sine rule, we will find 
\begin{equation}
    \phi = \arcsin{\left(\frac{s_1\sin{\frac{\theta}{2}}(1+\frac{s_2}{s_1})}{s}\right)}
    \label{eq:phi}. 
\end{equation}
Substituting equation~(\ref{eq:phi}) into the left hand side of equation~(\ref{eq:vv_equal}) and applying the trigonometric identities \(\cos{2\phi} = 1-2\sin{\phi}^2\) and \(\sin{\frac{\theta}{2}} = \frac{1-\cos{\theta}}{2}\), we will find 
\begin{equation}
    \mathrm{LHS} = - \frac{1}{2}\left(\frac{s^2(1-\cos{\theta})}{s^2}-\frac{(1-\cos{\theta})(s_1+s_2)^2}{s^2}\right) - \frac{2}{3}\cos{\theta}.
    \label{eq:LHS_simp}
\end{equation}
Notice the right hand side of equation~(\ref{eq:vv_equal}) also contains \(-\frac{2}{3}\cos{\theta}\). Therefore, we now only have to prove 
\begin{equation}
    \frac{1}{2}\left(\frac{s^2(1-\cos{\theta})}{s^2}-\frac{(1-\cos{\theta})(s_1+s_2)^2}{s^2}\right) = -\frac{s_1s_2}{s^2}(\sin{\theta})^2.
    \label{eq:vv_equal_simp}
\end{equation}
Substituting \((\sin{\frac{\theta}{2}})^2 = \frac{1-\cos{\theta}}{2}\), left hand side of equation~(\ref{eq:vv_equal_simp}) becomes 
\begin{equation}
    \mathrm{LHS} = (\sin{\frac{\theta}{2}})^2\left(1-\frac{(s_1+s_2)^2}{s^2}\right). 
    \label{eq:vv_LHS_simp}
\end{equation}
Using the double angle formula, the right-hand side of equation~(\ref{eq:vv_equal_simp}) becomes
\begin{equation}
    \mathrm{RHS} = -\frac{4s_1s_2}{s^2}(\sin{\frac{\theta}{2}})^2(\cos{\frac{\theta}{2}})^2.
    \label{eq:vv_RHS}
\end{equation}
\citet{Castorina_2018} states the length of the angular bisector \(d\) is given by 
\begin{align}
    d^2 &= s_1s_2\left[1-\frac{(\hat{s_1}-\hat{s_2})^2}{(s_1+s_2)^2}\right] \nonumber \\
    &= \frac{4s_1^2s_2^2}{(s_1+s_2)^2}(\cos{\frac{\theta}{2}})^2. 
    \label{eq:abt_d}
\end{align} 
Substituting in the second line of equation~(\ref{eq:abt_d}) to equation~(\ref{eq:vv_RHS}), we get 
\begin{equation}
    \mathrm{RHS} = -\frac{(s_1+s_2)^2d^2}{s_1s_2s^2}(\sin{\frac{\theta}{2}})^2.
    \label{eq:vv_RHS_simp}
\end{equation}
Then substituting in the first line of equation~(\ref{eq:abt_d}) for \(d^2\), we will find 
\begin{align}
    \mathrm{RHS} &= (\sin{\frac{\theta}{2}})^2\left[1-\frac{(s_1+s_2)^2}{s^2}\right] \nonumber \\
    &= \mathrm{LHS}. 
    \label{eq:equal_final}
\end{align}
Therefore, the formula for the velocity auto-covariance matrix in this paper is mathematically equivalent to the equations in \citet{Adams_2017} and \citet{ Ma_2011}. 

\section{Derivative of the logarithmic likelihood function with respect to the free parameter}
\label{sec:derivative}
The full covariance matrix is given by 
\begin{equation}
    \begin{pmatrix}
        \mat{C}_{gg}^{\rm err} + \mat{C}_{gg}^{\rm add} & \mat{C}_{g \eta}^{\rm grid} \\
        \mat{C}_{\eta g}^{\rm grid} & \mat{C}_{\eta \eta}^{\rm err}.
    \end{pmatrix}
    \label{eq:full_cov}
\end{equation}
We can rewrite the full covariance matrix as
\begin{equation}
\begin{split}
    \mat{C} = \sum_{p,q}^{l_{\rm max}} \sigma_g^{2(p+q)}
    \Bigg[(f\sigma_8)^2
    \begin{pmatrix}
        \mat{C}_{gg,p,q}^{\theta \theta} & \mat{C}_{g\eta, p, q}^{\theta \theta} \\
        \mat{C}_{\eta g, p, q}^{\theta \theta} & \mat{C}_{\eta \eta, p, q}
    \end{pmatrix}
    + fb\sigma_8^2 \\
    \begin{pmatrix}
        \mat{C}_{gg. p, q}^{m \theta} & \mat{C}_{g\eta, p,q }^{m \theta} \\
        \mat{C}_{\eta g, p, q}^{m \theta} & \mat{0}_{\eta \eta} 
    \end{pmatrix}
    +(b\sigma_8)^2
    \begin{pmatrix}
        \mat{C}_{gg, p, q}^{mm} & \mat{0}_{g\eta} \\
        \mat{0}_{\eta g} & \mat{0}_{\eta \eta}
    \end{pmatrix}
    +(b_{add}\sigma_8)^2 \\
    \begin{pmatrix}
        \mat{C}_{gg, p, q}^{\rm badd} & \mat{0}_{g \eta}\\
        \mat{0}_{\eta g} & \mat{0}_{\eta \eta}
    \end{pmatrix}
     \Bigg]
     + \sigma_v^2 
    \begin{pmatrix}
        \mat{0}_{gg} & \mat{0}_{g \eta} \\
        \mat{0}_{\eta g} & \eta_{vv}^{\rm grid}
    \end{pmatrix}
    +
    \begin{pmatrix}
        \mat{C}_{gg}^{sn} & \mat{0}_{g \eta} \\
        \mat{0}_{\eta g} & \mat{C}_{\eta \eta}^{\rm obs}.
    \end{pmatrix}
\end{split}
\label{eq:full_cov_split}
\end{equation}
Here \(\mat{C}_{xy}^{ab}\) denotes the component of the \(xy\) covariance matrix with the power spectrum \(P_{ab}\). Additionally, \(\mat{0}_{xy}\) is the zero matrix with the same dimension as the \(xy\) covariance matrix, \(\eta_{vv}^{\rm grid}\) is the matrix that converts the peculiar velocity to log-distance ratio for each grid cell, \(\mat{C}_{gg}^{sn}\) is the matrix that contains the shot noise of each grid cell, and \(\mat{C}_{\eta \eta}^{\rm obs}\) is the matrix contains the observational error of peculiar velocity at each grid cell. The summation stops at \(l_{\rm max}\) which depends on the highest order of the Taylor expansion. Let \(n_{\rm max}\) denotes the highest order of the Taylor expansion, then \(l_{max} = 4 (n_{\rm max}+1)\) for the galaxy auto-covariance matrix. For the cross-covariance matrix, the highest order of \(\sigma_g\) is \(2 (n_{\rm max}+1)\). For \(p+q > 2 (n_{\rm max}+1)\), the cross-covariance matrices are zero. For the velocity auto-covariance matrix, the highest order is zero, so for \(p+q > 0\), the velocity auto-covariance matrix is zero.

\section{The derivative of the logarithmic likelihood function with respect to the covariance matrix.}
\label{sec:second_derivative}
\subsection{First derivative}
The first derivative of the zero-point part of the logarithmic likelihood function with respect to the covariance matrix \citep{Petersen_2008}  
\begin{equation}
    \begin{split}
        \frac{d\ln{\boldsymbol{P(S|m)}}_{\rm ZP}}{d\mat{C(m)}} = -\frac{1}{2}\left(\frac{\sigma_y^2 \mat{C(m)}^{-1}\boldsymbol{x}\boldsymbol{x}^{T}\mat{C(m)}^{-1}}{N_x^2\sigma_y^2}\right. + \\
        \left. \frac{-2N_y\mat{C(m)}^{-1}\boldsymbol{S}\boldsymbol{x}^{T}\mat{C(m)}^{-1}N_x^2 + N_y^2\mat{C(m)}^{-1}\boldsymbol{x}\boldsymbol{x}^{T}\mat{C(m)}^{-1}}{N_x^4}\right).
    \end{split}
    \label{eq:deriv_zero_point}
\end{equation}

\subsection{Second derivative}
For the zero-point part of the logarithmic likelihood function, the second derivative of it with respect to the free parameters is given by 
\begin{align}
    \frac{d^2 \ln{\boldsymbol{P(S|m)}}_{\rm ZP}}{d\boldsymbol{m}^2} &= \frac{d \left(\frac{d\ln{\boldsymbol{P(S|m)}}_{\rm ZP}}{d\boldsymbol{m}}\right)}{d\boldsymbol{m}} \nonumber \\
    &= \frac{d \mathrm{Tr}\left(\frac{d\ln{\boldsymbol{P(S|m)}}_{\rm ZP}}{d\mat{C(m)}} \frac{d\mat{C(m)}}{d\boldsymbol{m}}\right)}{d\boldsymbol{m}} \nonumber \\
    &= \mathrm{Tr}\left(\frac{d \mathrm{Tr}\left(\frac{d\ln{\boldsymbol{P(S|m)}}_{\rm ZP}}{d\mat{C(m)}} \frac{d\mat{C(m)}}{d\boldsymbol{m}}\right)}{d\mat{C(m)}} \frac{d\mat{C(m)}}{d\boldsymbol{m}}\right). 
    \label{eq:dzero_point}
\end{align}
The only unknown term here is given by \(\frac{d \mathrm{Tr} \left(\frac{d\ln{\boldsymbol{P(S|m)}}_{\rm ZP}}{d\mat{C(m)}} \frac{d\mat{C(m)}}{d\boldsymbol{m}}\right)}{d\mat{C(m)}}\). Notice, \(\frac{d\mat{C(m)}}{d\boldsymbol{m}}\) is independent of \(\mat{C(m)}\) as shown in Appendix~\ref{sec:derivative}. For simplification, we will set \(\mat{M} = \frac{d\mat{C(m)}}{d\boldsymbol{m}}\) and separate the first derivative of the zero-point part of the logarithmic likelihood function (equation~\ref{eq:deriv_zero_point}) into two parts
\begin{align}
\begin{split}
     \frac{d \ln{\boldsymbol{P(S|m)}}_{\rm ZP}}{d\mat{C(m)}} = l_{d1} + l_{d2}
   = \left(-\frac{1}{2} \frac{\sigma_y^2 \mat{C(m)}^{-1}\boldsymbol{x}\boldsymbol{x}^{T}\mat{C(m)}^{-1}}{1+\boldsymbol{x}^{T}\mat{C(m)}^{-1}\boldsymbol{x}\sigma_y^2} \right) \\
   -\frac{1}{2N_x^4}\left(-2N_y\mat{C(m)}^{-1}\boldsymbol{S}\boldsymbol{x}^{T}\mat{C(m)}^{-1}N_x^2 + N_y^2\mat{C(m)}^{-1}\boldsymbol{x}\boldsymbol{x}^{T}\mat{C(m)}^{-1}\right). 
\end{split}
\end{align}
Using the online matrix calculus calculator\footnote{We used the matrix calculus calculator on this website \url{http://www.matrixcalculus.org/}.}, we get 
\begin{equation}
\begin{split}
     \frac{d \mathrm{Tr} \left(\frac{dl_{d1}}{d\mat{C(m)}} \bold{M}\right)}{d\mat{C(m)}} = -\left(\frac{\sigma_y^2}{N_x^2}x^{T}\boldsymbol{t_2}\boldsymbol{t_1}\boldsymbol{t_1}^T - \left(N_x^{-2}\boldsymbol{t_1}(\boldsymbol{t_1}^T \right. \right.\\
     \left. \left. \mat{M}^T\mat{C(m)}^{-1}) + N_x^{-2}\boldsymbol{t_2}\boldsymbol{t_1}^T\right)\right).
    \label{eq:d1}
\end{split}
\end{equation}
Here, we have \(\boldsymbol{t_1} = \mat{C(m)}^{-1}\boldsymbol{x}\) and \(\boldsymbol{t_2}=\mat{C(m)}^{-1}\mat{M}^T\boldsymbol{t_1}\). Similarly, we can find 
\begin{equation}
    \begin{split}
        \frac{d \mathrm{Tr} \left(\frac{dl_{d2}}{d\mat{C(m)}} \mat{M}\right)}{d\mat{C(m)}} = -\frac{1}{2}\left[ -\frac{2\boldsymbol{x}^{T}\boldsymbol{t_2}N_y^2}{N_x^6} + \frac{2 N_y}{N_x^4}\boldsymbol{x}^{T}\boldsymbol{t_2}\boldsymbol{t_1}\boldsymbol{t_{3}} \right.\\ \left. 
        + \mat{C(m)}^{-1}\boldsymbol{t_2}\boldsymbol{\boldsymbol{t_1}^T} +  N_x^4 N_y^2\boldsymbol{t_1}\boldsymbol{t_{4}}
         +\left(\frac{2 N_y}{N_x^4}\boldsymbol{S}^T \boldsymbol{t_2}\boldsymbol{t_1}\boldsymbol{t_1}^T-\right. \right.\\  
         \left. \left.\boldsymbol{S}^T \boldsymbol{t_2}\frac{2}{N_x^2}\boldsymbol{t_1}\boldsymbol{t_{3}}- \frac{2 N_y}{N_x^2}\boldsymbol{t_2}\boldsymbol{t_{3}}-\frac{2 N_y}{N_x^2}\boldsymbol{t_1}\boldsymbol{t_{5}}\right)\vphantom{\frac{1}{2}}\right],
    \end{split}
    \label{eq:d2}
\end{equation}
where \(\boldsymbol{t_{3}} = S^T \mat{C(m)}^{-1}, \boldsymbol{t_{4}} = \boldsymbol{t_1}^T\mat{M}^T\mat{C(m)}^{-1}\) and \(\boldsymbol{t_{5}} = \boldsymbol{t_{3}}\mat{M}^T\mat{C(m)}^{-1}\). Lastly, we can substitute equation~(\ref{eq:d1}) and equation~(\ref{eq:d2}) into equation~(\ref{eq:dzero_point}) to evaluate the second derivative of the zero-point part of the logarithmic likelihood with respect to the free parameters.

\bsp	
\label{lastpage}
\end{document}